\documentclass[aps,twocolumn,prd,showpacs,nofootinbib,showkeys,superscriptaddress,preprintnumbers]{revtex4-1}

\usepackage{graphicx,floatflt}
\usepackage{amsmath}
\usepackage{amstext}
\usepackage{epsfig}
\usepackage[usenames]{color}
\usepackage[dvipsnames]{xcolor}

\newcommand{\dd}{\mathrm{d}} 

\newcommand{\rr}{\mathrm}

\newcommand{\be}{\begin{equation}}
\newcommand{\ee}{\end{equation}}
\newcommand{\ba}{\begin{eqnarray}}
\newcommand{\ea}{\end{eqnarray}}

\begin{document}

\preprint{TTK-17-05 }

\title{Mimicking Dark Energy with the backreactions of gigaparsec inhomogeneities}

\author{S\'ebastien Clesse}
\email{clesse@physik.rwth-aachen.de}
\affiliation{Institute for Theoretical Particle Physics and Cosmology (TTK), RWTH Aachen University, D-52056 Aachen, Germany}

\author{Arnaud Roisin} \email{arnaud.roisin@student.unamur.be} 
\affiliation{Namur Center of Complex Systems (naXys), Department of Mathematics, University of Namur, Rempart de la Vierge 8, 5000 Namur, Belgium}

\author{Andr\'e F\"uzfa}
\email{andre.fuzfa@unamur.be}
\affiliation{Namur Center of Complex Systems (naXys), Department of Mathematics, University of Namur, Rempart de la Vierge 8, 5000 Namur, Belgium}
\affiliation{Centre for Cosmology, Particle Physics and Phenomenology,
Institute of Mathematics and Physics, Louvain University,
Chemin du Cyclotron 2, 1348 Louvain-la-Neuve, Belgium}

\date{\today}

\begin{abstract}
Spatial averaging and time evolving are non-commutative operations in General Relativity, which questions the reliability of the FLRW model.   The long standing issue of the importance of backreactions induced by cosmic inhomogeneities is addressed for a toy model assuming a peak in the primordial spectrum of density perturbations and a simple CDM cosmology.   The backreactions of initial Hubble-size inhomogeneities are determined in a fully relativistic framework, from a series of simulations using the BSSN formalism of numerical relativity.   In the FLRW picture, these backreactions can be effectively described by two so-called morphon scalar fields, one of them acting at late time like a tiny cosmological constant.   Initial density contrasts ranging from  $10^{-2}$ down to $10^{-4}$, on scales crossing the Hubble radius between $z\sim 45 $ and $z\sim 1000$ respectively, i.e. comoving gigaparsec scales, mimic a Dark Energy (DE) component that can reach $\Omega_{\rr{DE}} \approx 0.7$ when extrapolated until today.  A similar effect is not excluded for lower density contrasts but our results are then strongly contaminated by numerical noise and thus hardly reliable.  A potentially detectable signature of this scenario is a phantom-like equation of state $w< -1$, at redshifts $z\gtrsim 4$ for a density contrast of $10^{-2}$ initially, relaxing slowly to $w \approx -1$ today.  
This new class of scenarios would send the fine-tuning and coincidence issues of Dark energy back to the mechanism at the origin of the primordial power spectrum enhancement, possibly in the context of inflation.   

\end{abstract}
\maketitle

\section{Introduction}

Despite increasingly accurate observations, such as the ones of the cosmic microwave background (CMB) anisotropies, of the distribution of the large scale structures and of the type-Ia supernovae, the nature of the Dark Energy driving the recent acceleration of the cosmic expansion~\cite{Riess:1998cb} remains a major enigma of the standard cosmological model.   Most of the possible explanations investigated so far enter in one of the following categories:  First, a cosmological constant (CC), the simplest but rather unsatisfying explanation, suffering from the so-called \textit{fine-tuning} and \textit{coincidence problems};  second, a modification of the matter sector, e.g. through the introduction of a scalar field; third, a modification of the gravity sector, e.g. f(R) theories~\cite{Carroll:2003wy,DeFelice:2010aj,DeFelice:2010aj}; fourth, an effect of large inhomogeneities, e.g. if we live close to the center of a big void~\cite{Cusin:2016kqx}.   

Even if the fourth category does not require any modification of General Relativity (GR) neither a theory beyond the standard model of particle physics, most of the attention was given to theories of modified gravity or involving new matter (usually scalar) fields, with a strong interplay between them since any modification of the energy-momentum tensor can be interpreted at the cosmological level as a modification of the gravity sector, and inversely.   Besides explaining the cosmic acceleration, those models need to satisfy the very stringent local constraints on gravity, coming from laboratory experiments (see among others ~\cite{Brax:2016wjk,Burrage:2015lya,Elder:2016yxm,Schlogel:2015uea}), solar system (see e.g.~\cite{Bertotti:2003rm,Barreira:2015aea}), and from the growth of density fluctuations on cosmological scales.  To pass these constraints, one usually has to invoke some screening mechanism suppressing locally the modifications of gravity, such as the chameleon~\cite{Khoury:2003rn,Brax:2008hh}, Vainshtein~\cite{Vainshtein:1972sx,Babichev:2013usa} and K-mouflage~\cite{Babichev:2009ee,Brax:2014wla,Barreira:2014gwa} mechanisms.   

Belonging to the fourth category, backreactions from matter inhomogeneities can possibly lead to an apparent acceleration of the expansion, see e.g.~\cite{Buchert:1999pq,Rasanen:2003fy,Rasanen:2006zw,Rasanen:2006kp,Larena:2008be,Clarkson:2011zq,Buchert:2011yu,Rasanen:2011ki,Buchert:2011sx}.  According to the Buchert's theorem, spatial averaging and time evolving are not commutative operations in General Relativity~\cite{Buchert:1999er}.   Evolving some averaged quantity, such as the expansion rate or a density field, assuming homogeneity and isotropy as in the Friedmann--Lema\^itre--Robertson--Walker (FLRW) model, is not equivalent to evolving this quantity locally using the full Einstein equations and then averaging it in a Riemannian, covariant way.  
In particular, inhomogeneities induce a backreaction when interpreting observations in the FLRW picture, which can be effectively described by a minimally coupled scalar field, the so-called \textit{morphon} field~\cite{Buchert:2006ya}.  It has a non-restricted effective equation of state that can evolve from $w<-1$ to $w>-1$ (referred respectively as phantom-like and stiff matter fluids), and inversely, without implying any theoretical issue since the underlying theory is General Relativity and the morphon field has no physical existence.   As a result, the backreactions could mimic an accelerated expansion for an observer assuming homogeneity and isotropy, even if  locally the expansion rate decelerates everywhere.  

How important are the backreactions is a highly non trivial, and not yet entirely solved question (see e.g.~\cite{Buchert:2015iva} for a recent review and discussion), since their determination would require to solve at any point the full non-linear GR equations, over the whole cosmic history.  
The magnitude of the effect, and whether it can explain or not the observed cosmic acceleration, are still today controversial issues, even if the most recent studies tend to agree that the expected $10^{-5}$ primordial density fluctuations cannot induce a detectable level of late-time backreactions~\cite{Adamek:2014gva,Adamek:2015eda,Giblin:2015vwq,Bentivegna:2015flc} when the density field becomes non-linear on cosmological scales.   Several approaches have been considered to tackle the problem.  Among others, let us mention combined N-body simulations of dark matter with hydrodynamic simulations of the linear metric fluctuations~\cite{Adamek:2014gva,Adamek:2015eda};  swiss-cheese models~\cite{Biswas:2007gi,Valkenburg:2009iw,Lavinto:2013exa,Lavinto:2015iba} based on the Lema\^itre--Tolman--Bondi solution of the Einstein equations; and finally, since very recently, numerical relativity~\cite{Giblin:2015vwq,Bentivegna:2015flc,Bentivegna:2016stg}.   

Ultimately, the use of numerical relativity will be unavoidable in order to disregard the various possible approximations and to determine accurately and unambiguously the importance of the cosmological backreactions.  In this view, the development of  stable methods of numerical relativity in the context of astrophysical systems (black holes, neutron stars...), and particularly the Baumgarte--Shapiro--Shibata--Nakamura (BSSN) formalism~\cite{Shibata:1995we,Baumgarte:1998te}, combined with the never-ending growth of computational facilities, should allow to extend numerical relativity methods to various cosmological problems~\cite{East:2015ggf,Clough:2016ymm,Giblin:2015vwq,Bentivegna:2015flc,Bentivegna:2016stg,Rekier:2014rqa,Rekier:2015kxa,Rekier:2015isa,Braden:2016tjn}.  
Recently, the first backreaction studies based on the BSSN formalism~\cite{Giblin:2015vwq,Bentivegna:2015flc,Bentivegna:2016stg} have considered relatively simple initial density configurations.  Their main goal was to determine wether backreactions can reach a detectable level with future observations, provided a level of matter fluctuations on Hubble-scales as expected from CMB anisotropies.  

In this paper, we relax the assumption that dark matter fluctuations are initially at the $\sim 10^{-5}$ level and look at determining what is the required amplitude to get backreactions that could eventually mimic Dark Energy.   This approach is motivated since CMB and LSS only probe comoving modes within the range $10^{-4} - 10 \, \rr{Mpc}^{-1}$ and do not prevent a strong enhancement of power on larger or smaller scales.   Taking this point of view, the level of backreactions has been determined from lattice simulations in numerical relativity, and interpreted in terms of the density and equation of state of two apparent \textit{morphon} scalar fields.   Compared to~\cite{Buchert:2006ya} in which a single \textit{morphon} field was introduced, we have further considered the backreaction induced on the conservation equation of the energy-momentum tensor, which can be effectively described by a second \textit{morphon} field and which plays a crucial role here.  Distinguishing these two \textit{morphons} allows to determine which effect is able to reproduce a CC-like effect in the FLRW picture.  Finally, we have emphasized the possible ambiguity in the definition of the \textit{scale factor}, that can either be defined by how scales the total volume of the considered spatial domain, either as the averaged (in a Riemannian way) of the local scale factor, either related to the Riemannian averaging of the local scaling of proper lengths, the latter being the one related to observations through redshift and distance measurements.  

Our main result is that cold dark matter density contrasts in the range $\sim 10^{-4} - 10^{-2}$,  crossing initially the Hubble radius, induce backreactions acting like a tiny CC at late-times, in the FLRW picture.  When extrapolated to low redshifts, the associated \textit{morphon} field would finally dominate the energy density of the Universe and be able to mimic Dark Energy.   This happens for instance for initial $10^{-2}$ fluctuations at redshifts $z\sim 45 $, i.e. for comoving modes $k\sim 10^{-4}\, \rr{Mpc}^{-1}$ corresponding today to gigaparsec scales.    However this scenario should still be considered as a toy model, probably in some tension or even strongly disfavored by CMB anisotropy observations.   Nevertheless our results could open a whole new class of possible Dark Energy models, those exhibiting some peak in the primordial power spectrum of density fluctuations and inducing important backreactions.    Finally, we find that the time evolution of the backreactions is different from a cosmological constant at high redshifts.  More precisely they can be interpreted as a fluid with a phantom-like equation of state $w < -1$, an effect that could be probed by future LSS experiments like Euclid, Lyman-alpha forest, and 21-cm experiments like the Square Kilometre Array.  

Another goal of this paper is to pave the way of large simulations of structure formation using numerical relativity.  For this purpose, we have developed two independent codes based on the 3+1 BSSN formalism, for generic initial matter fluctuations with periodic boundary conditions.  The codes are gathered within the \textit{Inhomogeneous Cosmology And Relativistic Universe Simulations} (\texttt{ICARUS}) package, that will be soon made publicly available.   The codes i) solve the initial condition problem using a relaxation method to find the conformal factor of the metric respecting the Hamiltonian constraint equation,  ii) solve on a real-space 3D lattice, the full and non-linear evolution equations of general relativity in the synchronous gauge, using the BSSN method, and for periodic boundary conditions; iii) monitor the Hamiltonian constraint all during the free propagation scheme of numerical relativity, and perform some post-processing analysis like Riemannian averaging of the relevant quantities.    The development of two independent codes has allowed to cross-check our results at all the stages of this work.  

The paper is organized as follows.  In Section~\ref{sec:morphon} is set the correspondance between backreactions and the effects of two \textit{morphon} scalar fields in the FLRW picture.   In Section~\ref{sec:BSSN}, the evolution and constraint equations in the BSSN formalism of numerical relativity are given.   Our initial conditions are described in Section~\ref{sec:IC}.  After describing the numerical implementation of the BSSN equations and the code validation procedure in Section~\ref{sec:implementation}, the results of our simulations are presented in Section~\ref{sec:results}.   These are discussed in Section~\ref{sec:backreactions} with a particular focus on whether backreactions can mimic dark energy and lead to specific, potentially detectable signatures.   We summarize and discuss some interesting perspectives in Section~\ref{sec:ccl}. 

\section{Backreactions and the morphons} \label{sec:morphon}

Let us assume from know the simplest case of a pressureless cold dark matter Universe. 
The spatial average of some scalar quantity $\psi $ (such as the density, the scale factor,...) at time $t$, over some spatial compact domain $\mathcal D$ is given by
\be
\langle \psi \rangle (t)  = \frac{ \int_{\mathcal D} \dd^3 x \  \psi(t,{\bf x}) \sqrt{\gamma(t,{\bf x})}  }{\int_{\mathcal D} \dd^3 x \sqrt{\gamma(t,{\bf x})}}~,
\ee
with $\gamma$ the determinant of the projected 3-metric on the chosen spatial hypersurface.  In general, $\langle \psi \rangle (t) $ is different than $\bar \psi (t) $ obtained assuming homogeneity and isotropy prior to solve the dynamical Friedmann-Lema\^itre and conservation equations and get the evolution of $\bar \psi$. 

One can define the scale factor $a_\mathcal D (t)$ scaling the volume of the spatial domain like $a_\mathcal D^3(t)$.  It is in general not equivalent to the averaged local scale factor $\langle a(t,{\bf x}) \rangle \propto \langle [\sqrt{\gamma(t,{\bf x})}]^{1/3} \rangle$.   The evolution of $a_\mathcal D (t)$  is governed by~\cite{Buchert:2015iva}
\be  \label{eq:adotoaback}
\left(\frac{\dot a_{\mathcal D} }{a_{\mathcal D} } \right)^2 = \frac{8 \pi G}{3} \langle \rho \rangle_{\mathcal D} - \frac 1 6 ( Q_{\mathcal D} + \langle ^3R \rangle_{\mathcal D})~,
\ee
\be
\frac{\ddot a_{\mathcal D} }{ a_{\mathcal D}} = -\frac{4 \pi G}{3} \langle \rho \rangle_{\mathcal D}  + \frac 1 3 Q_{\mathcal D}~,
\ee
in which the kinematical backreaction variable $Q_{\mathcal D} (t) $ has been introduced, taking account for various effects of the local inhomogeneities, and where $^3 R$ is the local Ricci scalar of the 3-metric.  
These equations can be rewritten in a way reminiscent of the FLRW equations by introducing an \textit{apparent} scalar field $\varphi$, the so-called \textit{morphon} field~\cite{Buchert:2006ya}, with a density $\rho_\varphi$, a pressure $p_\varphi$ and an equation of state $w_\varphi$.   One obtains 
\be \label{eq:adotoa}
\left(\frac{\dot a_{\mathcal D} }{a_{\mathcal D} } \right)^2 =  \frac{8 \pi G}{3} (\langle \rho \rangle_{\mathcal D} + \rho_\varphi )~,
\ee
\be \label{eq:addotoa}
\frac{\ddot a_{\mathcal D} }{ a_{\mathcal D}} =  -\frac{4 \pi G}{3} (\langle \rho \rangle_{\mathcal D}  + \rho_\varphi + 3 p_\varphi )~,
\ee
so that both $\rho_\varphi$ and  $p_\varphi$ can be inferred from $a_{\mathcal D}(t)$ and its time derivatives, and from the averaged density $\langle \rho \rangle_{\mathcal D}$.  This sets the correspondance between the backreactions in the real Universe and the apparent scalar field in the FLRW picture that is used when interpreting the observations.   

The observations often involve redshift measurements, and redshifts are related to photon wave-vectors along null geodesics, $k^\mu = \dd x^\mu / \dd t $ where $t$ is  the cosmic time, so that in the 3+1 decomposition described in the next section, the wavelength of photons propagating in the \texttt{x} direction\footnote{On the lattice, we denote by $(\texttt{x, y, z})$ the spatial coordinates $(x^1, x^2, x^3)$.} scales with $\sqrt{ \gamma_{11}}$ ($\gamma_{ij}$ being the projected spatial metric).  This factor also scales proper lengths locally and it is involved in the calculation of comoving, diameter and luminosity distances.   It is therefore the scale factor directly inferred from the observations, and is denoted here $\alpha (x,t)$.  In an inhomogeneous Universe, $\langle \alpha \rangle (t)  \neq a_{\mathcal D}(t)$.   The difference between $a_{\mathcal D}$, $\langle a \rangle$ and $\langle \alpha \rangle$ is  subtle but of importance for determining how the backreactions affect the Universe's dynamics seen in the FLRW picture.    In order to determine them, one has to evaluate  $\rho_\varphi [\langle \alpha \rangle (t)]$ and  $p_\varphi [\langle \alpha \rangle (t)]$.   In this paper, we neglect the effects of inhomogeneous light propagation~\cite{Fleury:2015hgz,Brouzakis:2007zi,Umeh:2010pr} and assume that all points in the lattice simulations are at a fixed light-travel distance from a late-time observer.   Including this effect would require to implement a ray-tracing method, which is left for a future work.

In addition to the above mentioned effects, the density satisfies locally the energy-momentum tensor conservation equation, which in the synchronous gauge reads
\be
\partial_t  \rho(t,x)= - 3   \frac{ \dot a(t,\bf{x})}{a(t,\bf{x})} \rho(t,\bf{x})\,.
\ee
Because time-evolving does not commute with spatial averaging, one has $\langle \partial_t \rho \rangle_{\mathcal D} \neq \partial_t \langle \rho \rangle_{\mathcal D}$.  Thus the previous relation must be averaged out using
\ba
\langle \partial_t \rho \rangle_{\mathcal D} & = &  \langle - 3   \frac{ \dot a(t,{\bf x})}{a(t,{\bf x})} \rho(t,{\bf x}) \rangle \\
& = & - 3 \left[  \frac{\dot a_{\mathcal D}}{a_{\mathcal D}} \langle \rho \rangle +  \left(   \langle  \frac{ \dot a(t,{\bf x})}{a(t,{\bf x})} \rho(t,{\bf x}) \rangle - \frac{\dot a_{\mathcal D}}{a_{\mathcal D}} \langle \rho \rangle  \right) \right] \\
& = & - 3  \frac{\dot a_{\mathcal D}}{a_{\mathcal D}} \langle \rho \rangle + Q_\rho \, ,
\ea
where we have introduced a new backreaction term $Q_\rho$ that was not considered in~\cite{Buchert:2006ya} as well as other works studying the backreactions using numerical relativity~\cite{Giblin:2015vwq,Bentivegna:2015flc,Bentivegna:2016stg}.   $Q_\rho$ is non-zero in general and one can interpret this term in the FLRW picture as the presence of a second \textit{morphon} field $\chi$ with a density $\rho_\chi $ and a time-dependent equation of state $w_\chi$,
\be \label{eq:rhochi}
\langle \rho \rangle_{\mathcal D} (t) = \langle \rho\rangle^{\rr{ini}} \langle \alpha \rangle^{-3} + \rho_\chi  (\langle \alpha \rangle) ~, 
\ee
\be \label{eq:wchi}
w_\chi = -1- \frac 1 3 \frac{\dd \ln \rho_\chi}{\dd \langle \alpha \rangle}~,
\ee
such that one recovers $\rho_\chi \propto \langle \alpha  \rangle^{-3(1+w)}$ for a constant $w_\chi$.   As it will be shown later, the backreactions of $\chi$ on the averaged density are crucial and leads to an equation of state reaching $w_\chi \approx-1 $.   Its tiny density becomes non-negligible with time and eventually becomes dominant and leads to an apparent Dark Energy component.  The dynamics of $\chi$ is also influenced by non-equivalence between $\langle \alpha \rangle (t) $ and $a_{\mathcal D}(t)$.   
We will also show that $w_\varphi \approx -1/3$  and thus that $\varphi$ acts like a curvature fluid.



\section{The BSSN formalism \\ of numerical relativity}  \label{sec:BSSN}

The BSSN formalism~\cite{Shibata:1995we,Baumgarte:1998te} uses the 3+1 decomposition of the metric (in geometrical units where $c=G=1$)
\be
\dd s^2 = (- \alpha^2 + \gamma_{lk} \beta^l \beta^k ) \dd t^2 + 2 \beta_i \dd t \dd x^i + \gamma_{ij} \dd x^i \dd x^j.
\ee

We have worked in the synchronous gauge in which the \textit{lapse}  and the \textit{shift} are respectively $\alpha = 1$ and $\beta^i = 0$.   
The 3-metric $\gamma_{ij}$ is then decomposed into a conformally related metric $\bar \gamma_{ij}$ of determinant $\bar \gamma = 1$,
\be
\gamma_{ij} = \rr e^{4 \phi } \bar \gamma_{ij}~,
\ee
where $\phi$ is the so-called conformal factor.   One also decomposes the extrinsic curvature tensor $K_{ij}$ into its trace $K$ and its conformally rescaled, traceless part $\bar A_{ij}$, 
\be
K_{ij} = \rr e^{4 \phi} \bar A_{ij} + \frac 1 3 \gamma_{ij} K.
\ee   
We focus on the simplified case where the Universe is filled only with a pressure-less fluid (that could be the Dark Matter plus eventually non-interacting baryons).  In the synchronous gauge, the only non-vanishing component of the energy-momentum tensor is the energy density $\rho$.   The Einstein's equations are equivalent to a set of evolution equations for the dynamical quantities $\phi, K, \bar \gamma_{ij}, \bar A_{ij}$, 
\ba
\partial_t \phi & = & - \frac 1 6 K   \label{eq:phievol}   \\
\partial_t K & = & \bar A_{ij} \bar A^{ij} + \frac 1 3 K^2 + 4 \pi \rho  \label{eq:Kevol}\\
\partial_t \bar \gamma_{ij} & = & - 2 \bar A_{ij} \\
\partial_t \bar A_{ij} & = & \rr e^{-4 \phi} R_{ij}^{\rr{TF}} + K \bar A_{ij} - 2 \bar \gamma^{kl} \bar A_{il} \bar A_{kj}~,  \label{eq:Aijevol}
\ea
where $R_{ij}^{\rr{TF}} \equiv R_{ij} - \gamma_{ij} R/3$ is the trace-free Ricci tensor of the 3-metric $\gamma_{ij}$.  In the BSSN formalism, in order to improve the numerical stability of the PDEs system, especially in the context of problems with singularities, one also defines the conformal connection functions $\bar  \Gamma ^i \equiv - \partial_j \bar \gamma^{ij}$, which are treated as dynamical variables obeying to their own evolution equations,
\be
\partial_t \bar \Gamma^i = 2 \bar \Gamma^i_{†j k } \bar A^{jk} - \frac 4 3 \bar \gamma^{ij} \partial_j K + 12 \bar A^{ij} \partial_j \phi~.
\ee
The last equation of evolution is the energy-momentum conservation equation, for the energy density,
\be \label{eq:rhoevol}
\partial_t \rho =  K \rho~.
\ee
In addition, the system needs to satisfy four constraint equations.  The first one is the Hamiltonian constraint,
\ba \label{eq:Hconstrain}
\mathcal H & = & \bar \gamma^{ij} \bar D_i \bar D_j \rr e^\phi - \frac{\rr e^\phi}{8} \bar R \nonumber \\
& &  + \frac{\rr e^{5 \phi}}{8} \bar A_{ij} \bar A^{ij} - \frac{\rr e^{5 \phi} }{12 } K^2 + 2 \pi \rr e^{5 \phi} \rho =  0,
\ea
where $\bar D_i $ denotes the covariant spatial derivative for the 3-metric $\bar \gamma_{ij}$.    The three other ones are the momentum constraints
\be \label{eq:momconstrain}
\bar D_j( \rr e^{6 \phi} \bar A^{ij} ) - \frac{2}{3} \rr e^{6 \phi} \bar \gamma^{ij} \bar D_{j} K = 0~.
\ee

Note that in the homogeneous case, the scale factor of the Universe is identified to $a = \rr e^{2 \phi}$, and the Hubble expansion rate $H \equiv \dot a / a = - K/3 $.  One recovers the usual Friedmann-Lema\^itre equations from the Hamiltonian constraint and the evolution equations (\ref{eq:Kevol}) and (\ref{eq:phievol}), as well as the energy conservation equation $\dot \rho = - 3 H \rho $ from Eq.~(\ref{eq:rhoevol}).

\section{Initial Conditions}  \label{sec:IC}

The initial conditions must satisfy the constraint equations (\ref{eq:Hconstrain}) and (\ref{eq:momconstrain}).   The second one is trivially satisfied if the initial hypersurface is chosen such that $\bar A_{ij} =0$ and if $ K$ and  $\bar \gamma_{ij}$ are constant everywhere on the lattice, which has been our simplifying assumption.  In addition we set $\bar \gamma_{ij} = \delta_{ij} $, so that the Hamiltonian constraint becomes
\be
\bar D_i \bar D_i \rr e^{\phi} = \frac 1 {12} \rr e^{5 \phi} \left(  K^2 -24 \pi \rho \right)~.
\ee
Because the initial conditions are fixed at a stage where the Universe is close to FLRW, we simply set 
\be
K_{\rr {ini}}^2 = 24 \pi \bar \rho_{\rr {ini}} 
\ee
initially, where $\bar \rho_{\rr {ini}}$ is the averaged initial density.   

This simplified choice of initial conditions is identical to the one made in \cite{Giblin:2015vwq,Bentivegna:2015flc}.   It is not expected to reproduce very accurately the reality, but nevertheless is a good approximation since it corresponds to the expectations (for this toy Universe) of a perturbed FLRW Universe, at fist order in the linear theory of cosmological perturbations.   
In this way the momentum constraints are trivially satisfied.  This choice also significantly simplifies the Hamiltonian constraint that can then be solved using an iterative method for elliptic problems, in order to determine the value of $\phi$ on the initial hypersurface, given the initial density $ \rho_{\rr{ini}}$.   This approach is similar to the one of~\cite{Bentivegna:2015flc} and is  opposite to the one of~\cite{Giblin:2015vwq} in which the density field was determined from a given pattern of $\phi$.  

We follow the same choice than~\cite{Bentivegna:2015flc} for the initial density $\rho_{\rr{ini}}$, with periodic fluctuations around the homogeneous value $ \bar \rho_{\rr{ini}}$  characterized by a series of modes,
\be \label{eq:rhoinit}
\rho ({\bf x} ) = \bar \rho \left\{1+ \sum_{k_x,k_y,k_z}  \tilde \delta({\bf k}) \cos[{\bf k} . {\bf x} + \theta({\bf k}) ] \right\}~.
\ee  
The phases $\theta({\bf k}) $ are taken to vanish in our simulations with a single wavelength mode along each spatial direction, and are set to a random value in the others.   The mode amplitudes $\tilde \delta ({\bf k})$ are set to values lower than, or eventually approaching, the limit of the non-linear regime.   
The number of modes is also limited by the size of the simulation, and was restricted to $k_{x,y,z} \leq 2 $ in each spatial direction, i.e. a total of 26 modes. 

We have fixed $\bar \rho_{\rr{ini}} = 1$ initially, which fixes the length and time units of our simulations  to $u_l = c \, u_t = \sqrt{8 \pi / 3} / H_{\rr {ini}} $ and relates them to the initial Hubble radius $H_{\rr {ini}}^{-1}$.

\begin{figure}
\begin{center}
\includegraphics[scale=0.8]{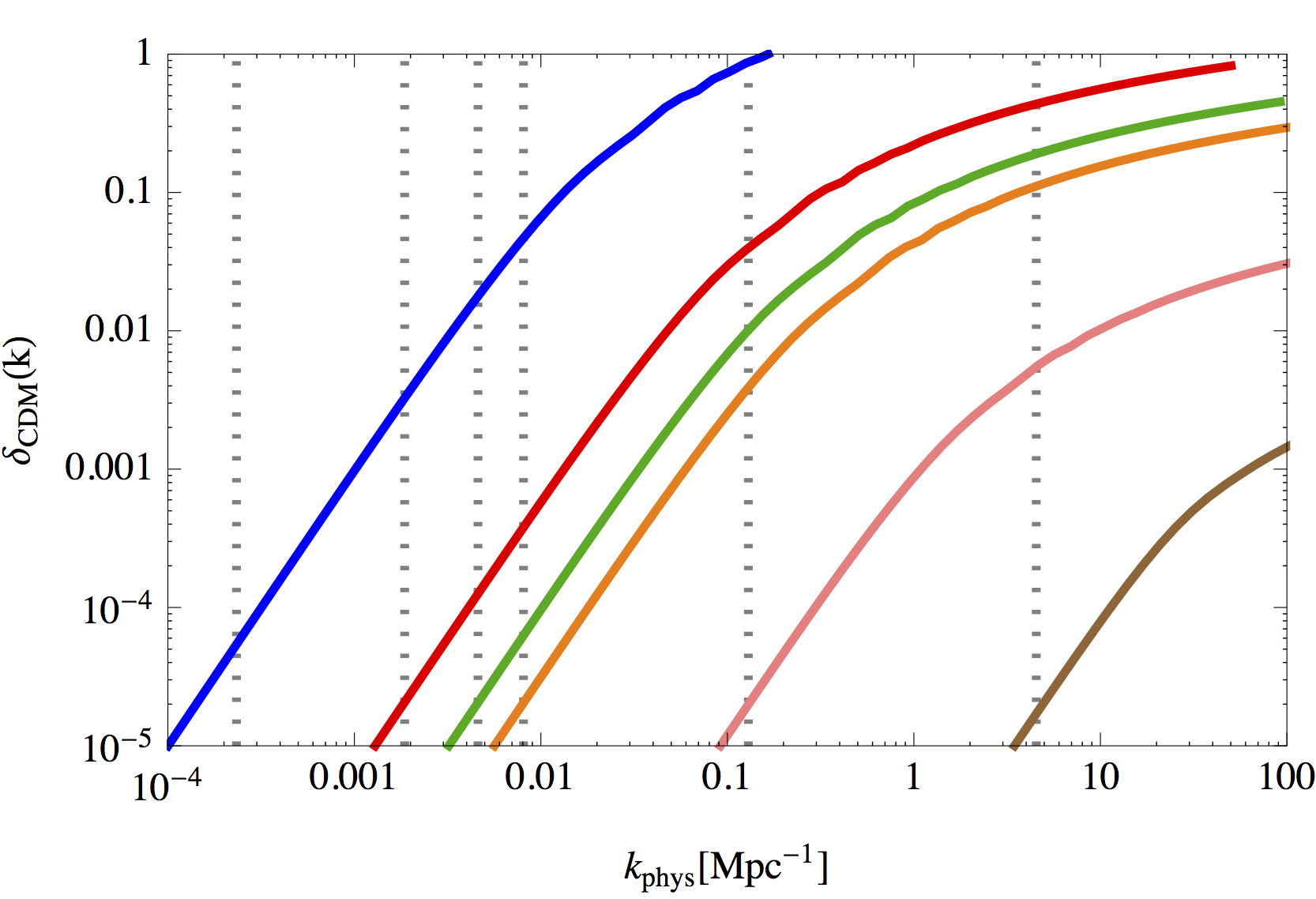}
\caption{Spectrum of CDM density fluctuations for the best-fit $\Lambda$-CDM cosmology, at redshifts $z=0,5,10,15, 100, 1000$ (respectively blue, red, green, orange, pink and brown), including non-linear corrections from \texttt{HALOFIT}~\cite{Smith:2002dz}.  Vertical dotted lines represent the corresponding value of $H(z)$. }
\label{fig:deltasLCDM}
\end{center}
\end{figure}

In order to ease the comparison with the $\Lambda$-CDM model, the spectrum of CDM fluctuations has been computed using 
 the \texttt{CAMB} code~\cite{Lewis:1999bs} at several redshifts, for the Planck best fit values~\cite{Ade:2015xua} of the six standard cosmological parameters $\{\Omega_b h^2, \Omega_c h^2, \theta, \tau, A_s, n_s \}$.   
 These spectra are displayed on Fig.~\ref{fig:deltasLCDM} as a function of physical wavenumbers, as well as the corresponding values of the Hubble rate.  
 
\section{Numerical implementation}  \label{sec:implementation}

\subsection{The \texttt{ICARUS} codes}
In order to solve the initial condition problem and the equations of evolutions in the BSSN formalism, on a real-space uniform lattice with periodic boundaries, we have developed and used a package called \textit{Inhomogeneous Cosmology And Relativistic Universe Simulations}, referred as \texttt{ICARUS}.  More details about the \texttt{ICARUS} codes will be provided soon in a dedicated paper, hereafter we present their principal features.   The package contains two codes, developed independently (one in \texttt{c} and one in \texttt{fortran}).   The BSSN equations are solved in the synchronous gauge, but in the future additional gauges could be included.  The use of periodic boundary conditions is well suited for cosmological applications\footnote{A possible alternative would be to implement Sommerfeld-type boundary conditions, as in Refs.~\cite{Rekier:2014rqa,Rekier:2015isa,Rekier:2015isa}.}, so that the lattice represents a patch of the Universe, reproduced infinitely along the three spatial dimensions.   

Regarding the initial conditions, as mentioned in the previous section, the codes solve first the Hamiltonian constraint Eq.~(\ref{eq:Hconstrain})  assuming $\bar A_{ij} = 0$ and $\bar \gamma = 1 $ initially.  For this purpose a simple Jacobi relaxation method in the real space has been implemented, and was found to be more accurate than an iterative method in the Fourier space (which works only for matter inhomogeneities expressed as a sum of wavelength modes, as in Eq.~(\ref{eq:rhoinit}), whereas a real-space method is more general).

Once the initial conditions of $\phi$ respecting the Hamiltonian constraint are found, the system formed by the BSSN evolution equations is solved on the lattice (the so-called free propagation scheme of numerical general relativity), by using a Heun's PE(CE)$^3$ predictor-corrector build on the Euler's explicit method.    The hamiltonian constraint is monitored during all the numerical integration.   Spatial derivatives are computed using fourth-order central differences schemes.   The codes allow to export any dynamical quantity at any time, for specific locations, for one-dimensional or two-dimensional spatial slices, or even for the full lattice.   Riemannian averaging can be performed at all time steps and for any dynamical and local variable such as the scale factors $a$ and $\alpha$, the matter density $\rho$, the extrinsic curvature $K$ related to the local expansion rate.   Averaging is then used to determine the effect of the backreactions in the FLRW picture, through the density $\rho_{\varphi,\chi}$, the pressure $p_{\varphi,\chi}$ and the equation of state $w_{\varphi,\chi}$ of the two \textit{morphon} fields.   These quantities are computed from the evolution of $\langle \rho \rangle $ as a function of $\langle \alpha \rangle $ using Eqs.~(\ref{eq:rhochi}) and~(\ref{eq:wchi}), and from the evolution of $a_\mathcal D$ and its time derivatives, using Eqs.~(\ref{eq:adotoa}) and~(\ref{eq:addotoa}).

For the present paper, we have run a series of simulations for one-dimensional, two-dimensional and three-dimensional matter inhomogeneities with lattice sizes up to $10^5$ points along the \texttt{x} direction in the 1D case  (note that a minimum of 3 points are also needed along \texttt{y}  and \texttt{z} directions for the computation of numerical derivatives with periodic boundary conditions), $1024^2$ in the 2D case and up to $128^3$ in the 3D case.  Nevertheless, for improving the long-time stability of the code, it is often more convenient to reduce the lattice size together with reducing the time step down to $c \Delta t \lesssim 10^{-4} \Delta x$, with a lattice physical length of the order of the initial Hubble radius.    For long runs, our codes have been parallelized using \texttt{openmp}.  The simulation proceeds up to the time where the Hamiltonian constraint starts to evolve exponentially, i.e. when the solution cannot be trusted anymore.

\begin{figure}
\begin{center}
\includegraphics[scale=1.1]{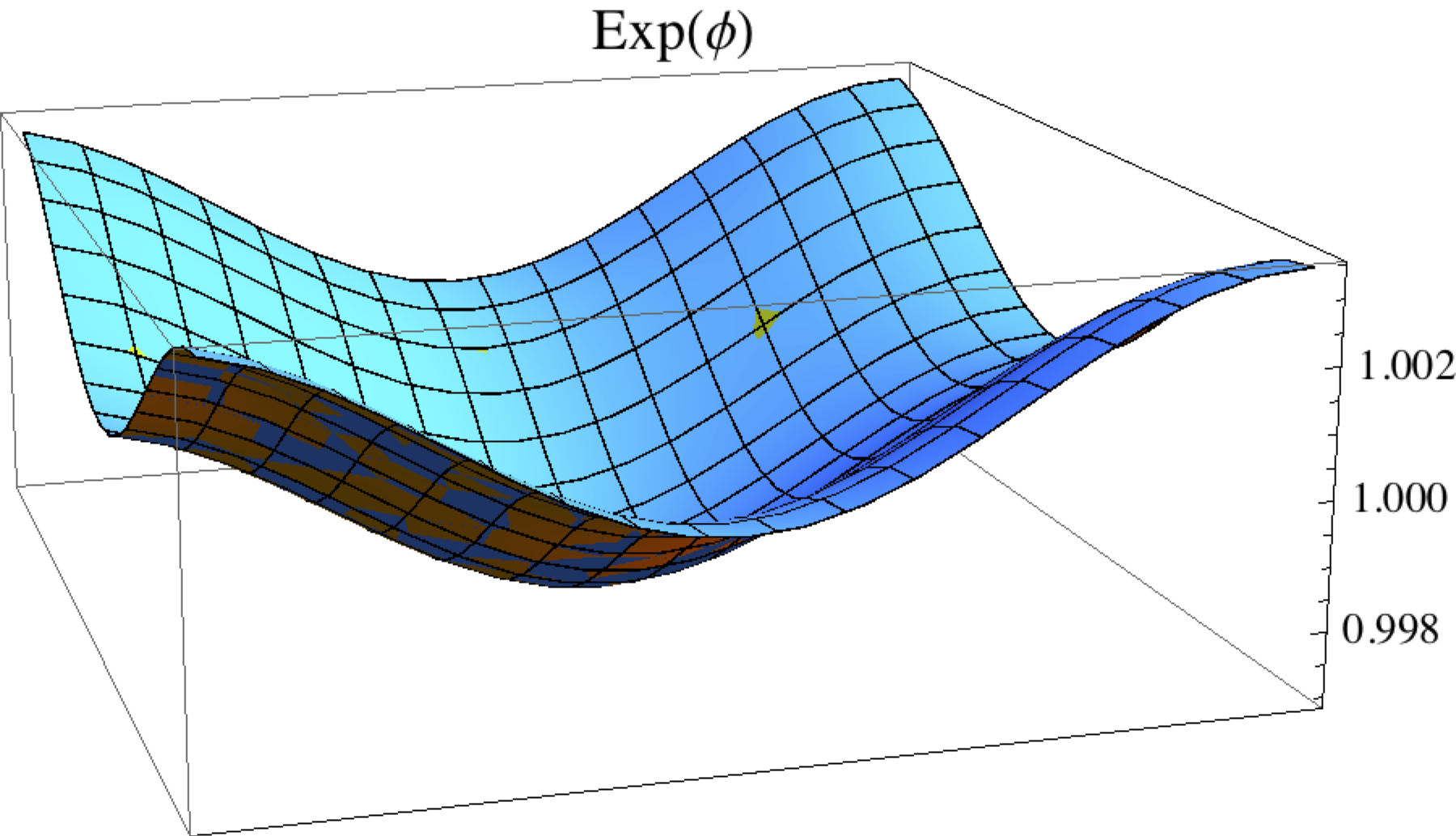} \\
\includegraphics[scale=1.1]{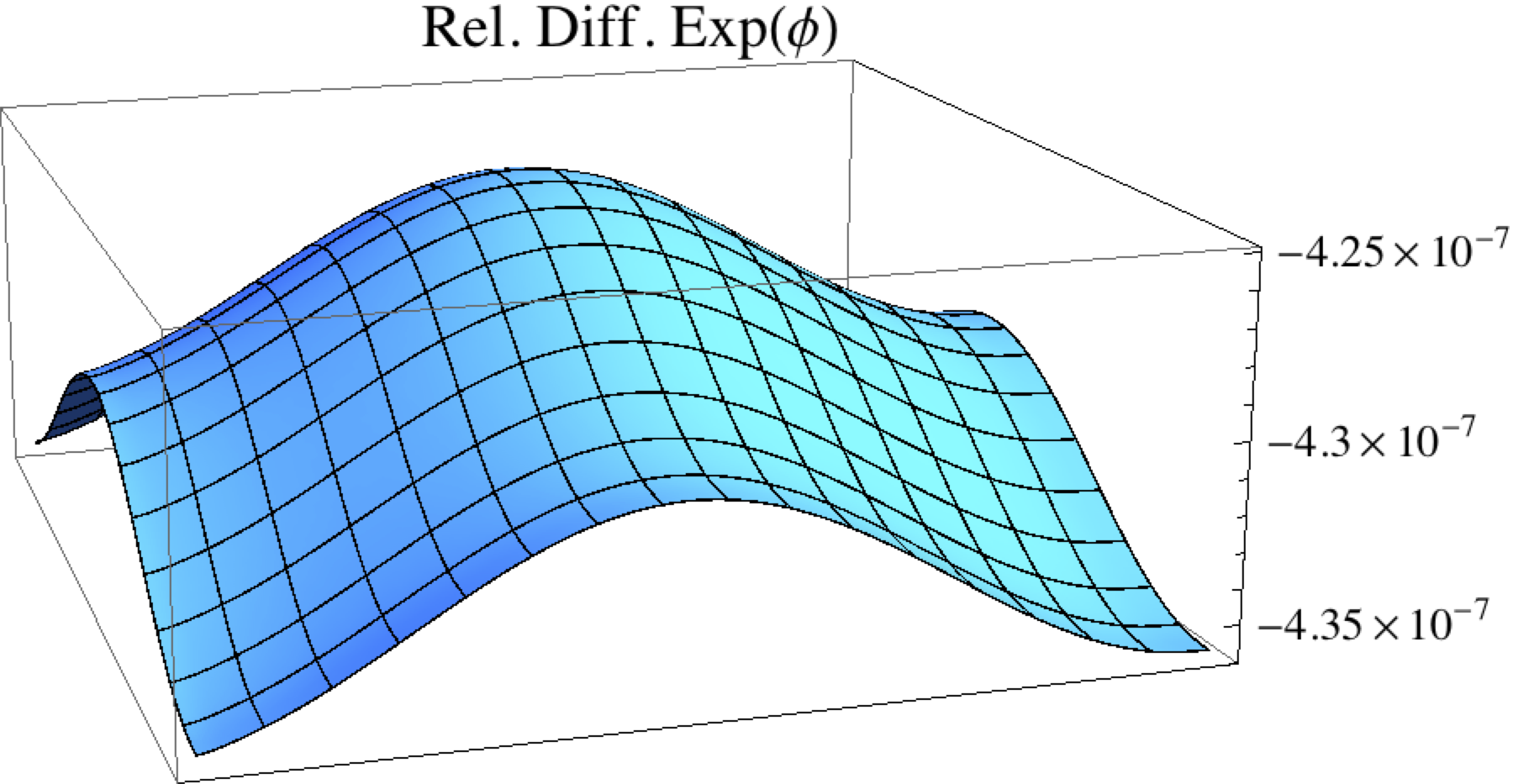} \\
\includegraphics[scale=1.1]{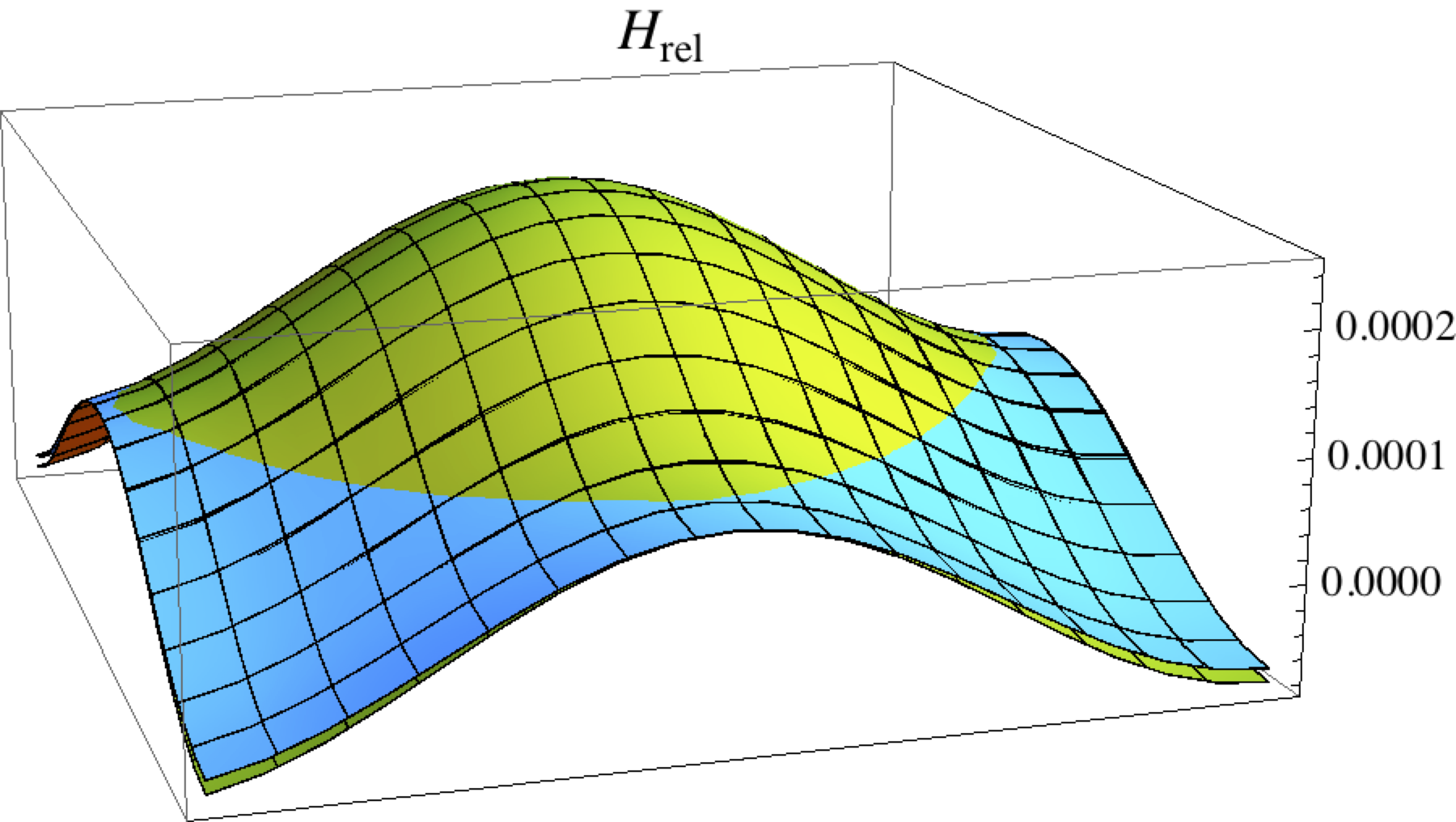} \\
\caption{Initial conformal factor $\exp(\phi)$ (top panel) and the relative difference between the two \texttt{ICARUS} codes (middle panel), for a two-dimensional inhomogeneity, a $20\times20\times3$ lattice of length $L = H_{\rr{ini}}^{-1}$ and a density contrast $\tilde \delta(k) =0.01$ with one single mode $k_{x,y} = 2 \pi / L$ along the \texttt{x} and \texttt{y} directions.  Bottom panel:  Hamiltonian constraint $\mathcal H^{\rr{rel.}} $ initially.     }
\label{fig:phiinitHrel}
\end{center}
\end{figure}

\subsection{Validation procedure}

\begin{figure}
\begin{center}
\includegraphics[scale=0.75]{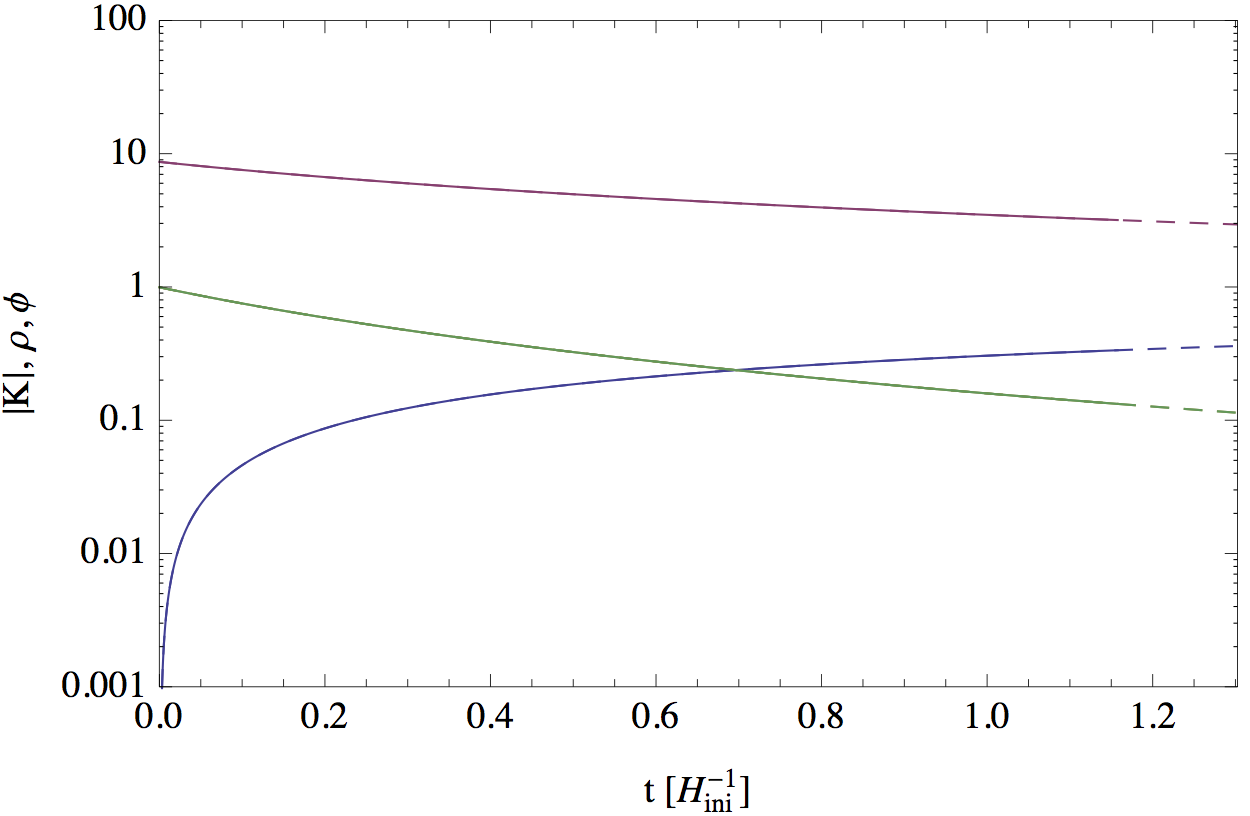} \\
\includegraphics[scale=0.75]{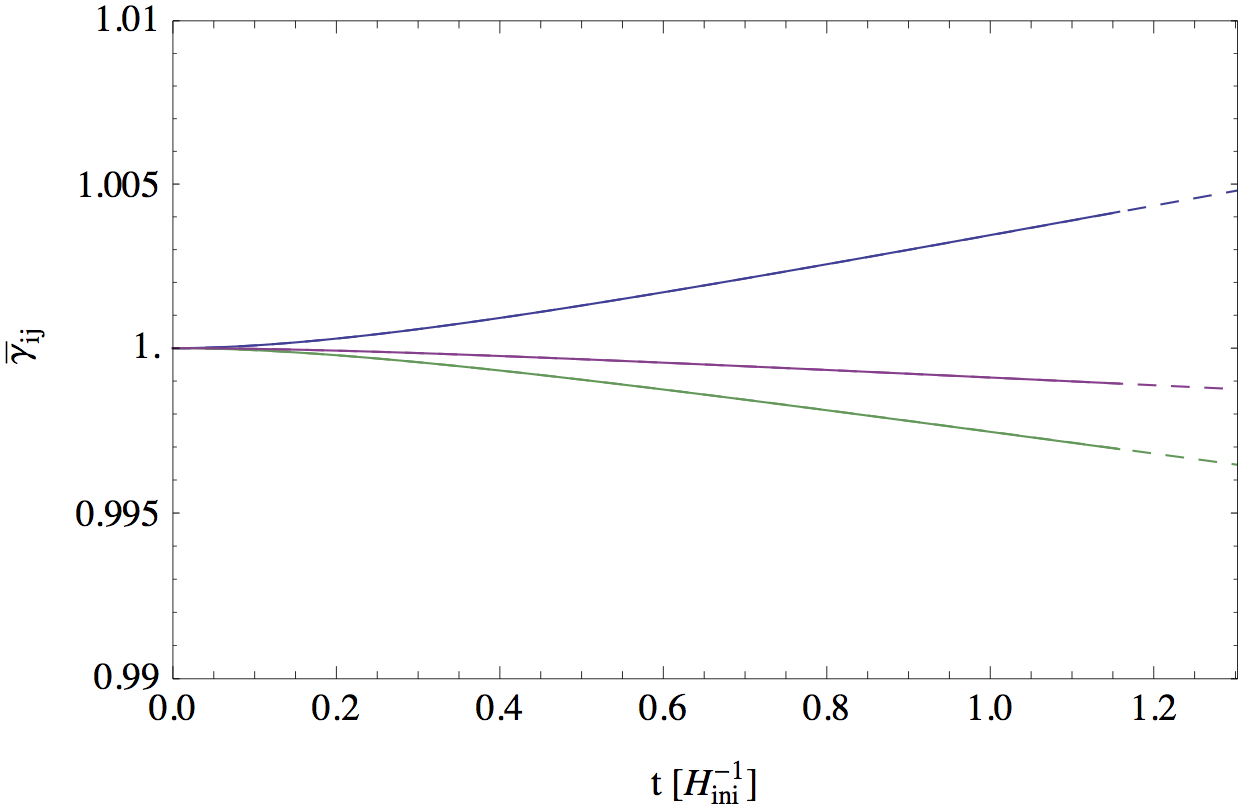} \\
\includegraphics[scale=0.75]{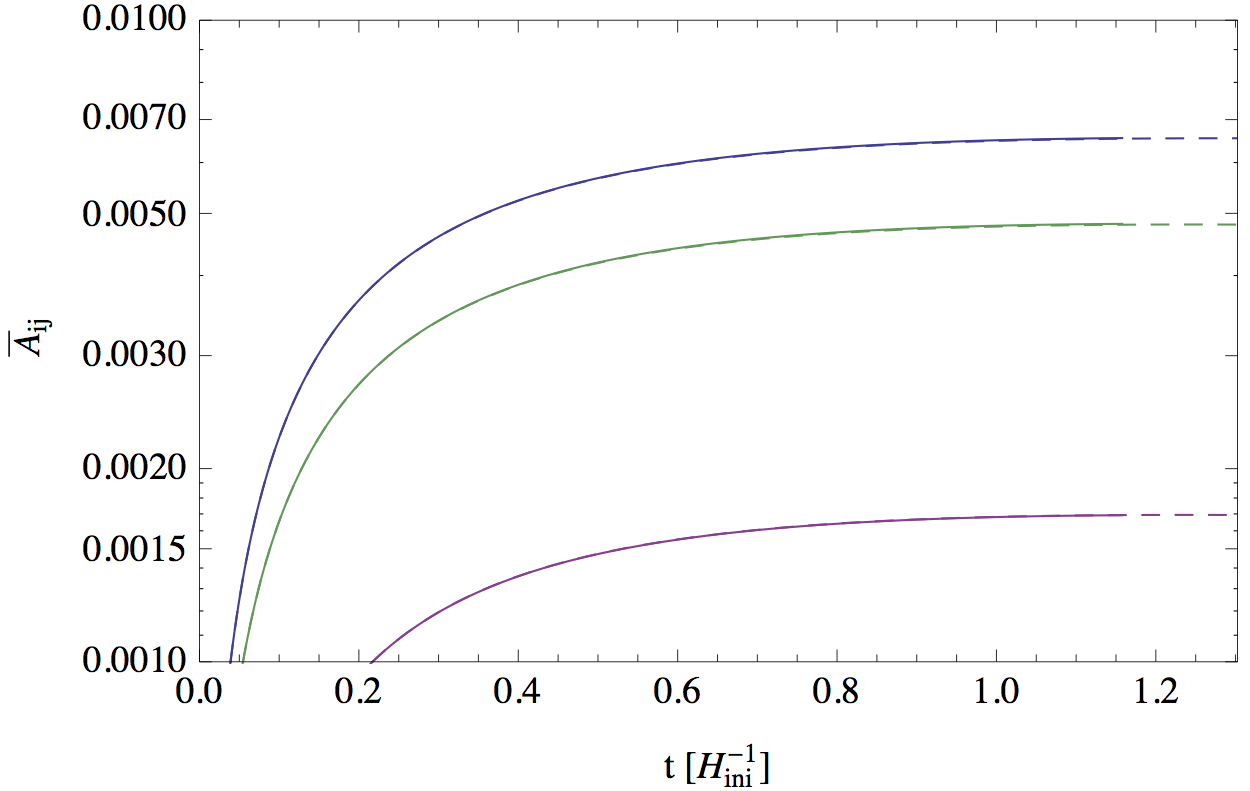} \\
\includegraphics[scale=0.75]{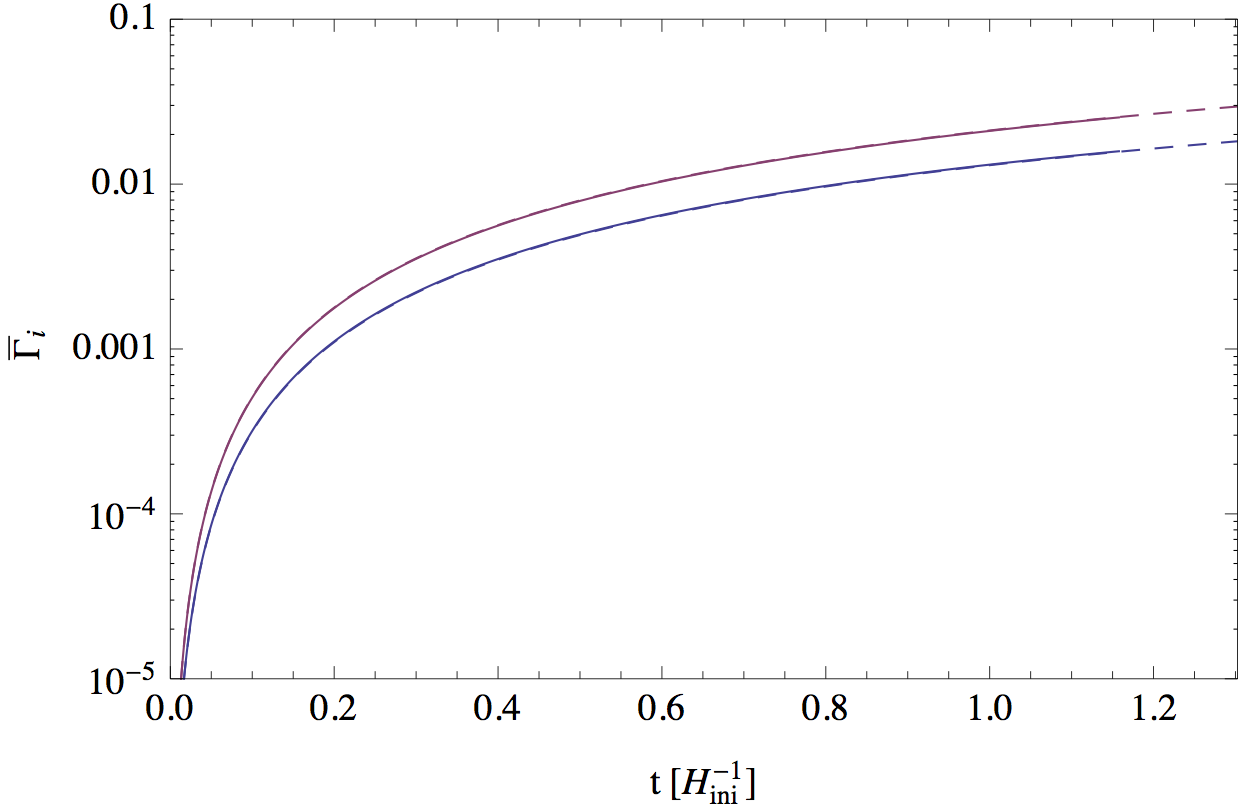} \\
\caption{One-point evolution of dynamical variables $\rho$ (top panel, green), $K$ (red), $\phi$ (blue), $\bar \gamma_{11} $ (second panel, blue), $\bar \gamma_{22} $ (red), $\bar \gamma_{33} $ (green), $\bar A_{11}$ (third panel, blue), $\bar A_{22}$ (red), $\bar A_{33}$ (green) and $\bar \Gamma_{1,2}$ (bottom panel, blue and red respectively), obtained with the two \texttt{ICARUS} code-1 (solid) and code-2 (dashed, superimposed), for initial conditions as in Fig.~\ref{fig:phiinitHrel}.  The time step is $\Delta t = 5 \times 10^{-5} H^{-1}_{\rr i}$.}
\label{fig:evolvalid}
\end{center}
\end{figure}

\begin{figure}
\begin{center}
\includegraphics[scale=0.8]{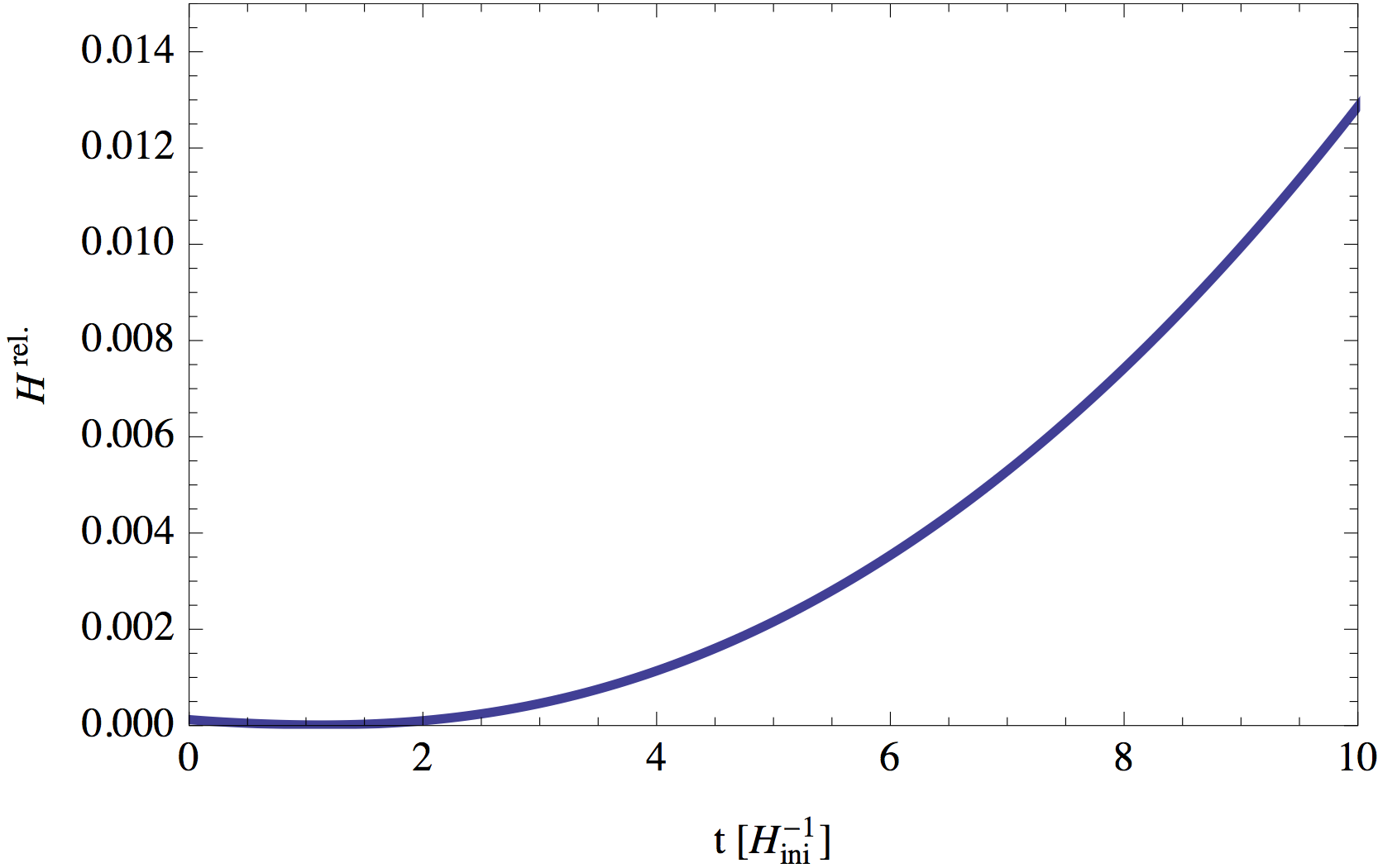}
\caption{Evolution of the Hamiltonian constraint $\mathcal H^{\rr{rel}}$, for initial conditions as in Fig.~\ref{fig:phiinitHrel}, and time step is $\Delta t = 5 \times 10^{-5} H^{-1}_{\rr i}$.}
\label{fig:Hrel}
\end{center}
\end{figure}

\begin{figure*}
\begin{center}
\includegraphics[scale=0.75]{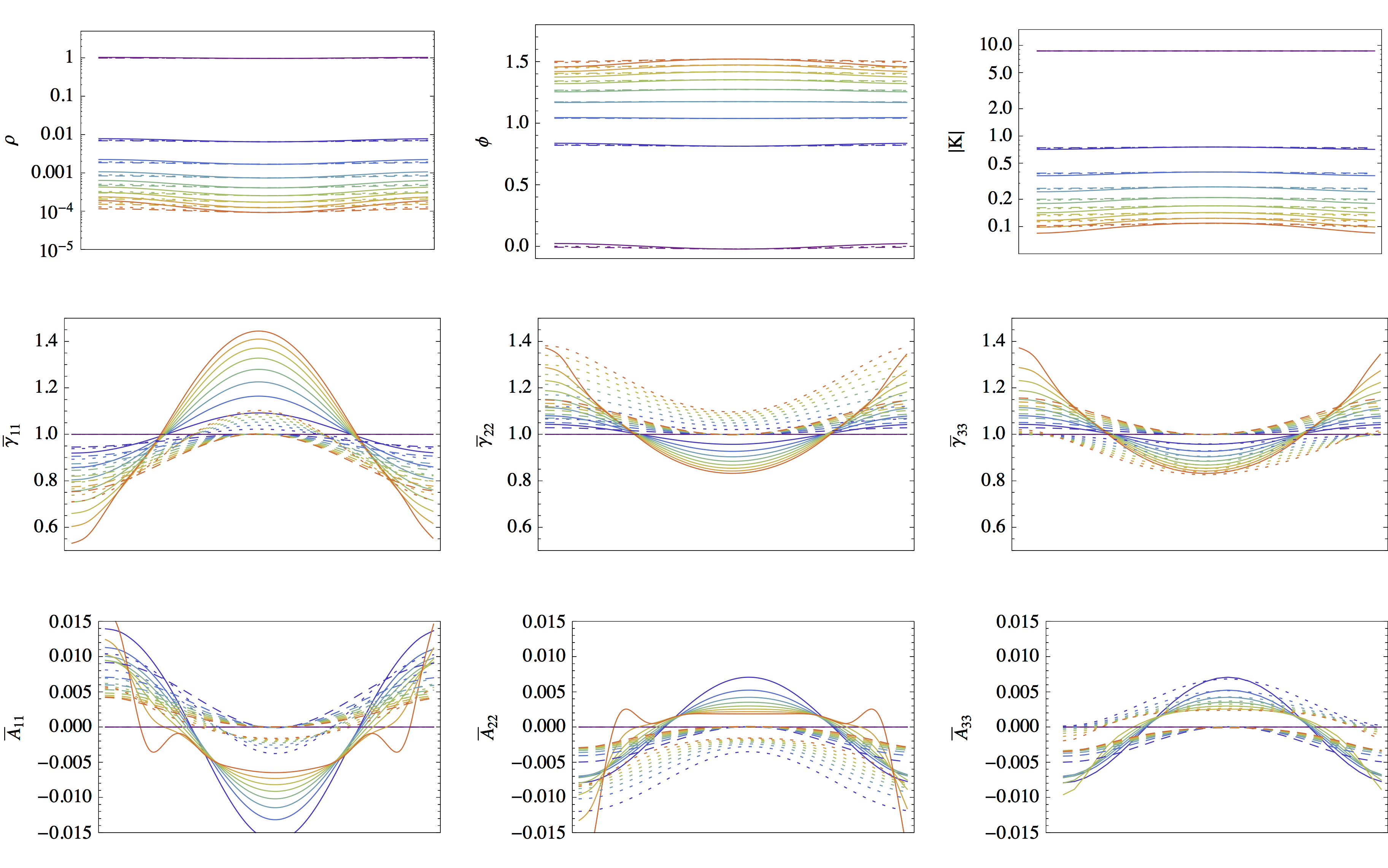} \\
\caption{One-dimensional lattice slice showing the time evolution (from blue to red when time increases) of the dynamical variables $\rho$ (top left), $K$ (top center), $\phi$ (top right), $\bar \gamma_{ii} $ (middle panels) and $\bar A_{ii}$ (bottom panels) for an initial one-dimensional inhomogeneity (single mode) with $\delta(k) = 0.03$ (solid lines), a two-dimensional inhomogeneity with $\delta(k) = 0.015$ (dotted lines), and a three-dimensional one with $\delta(k) = 0.01$ (dashed lines),  and $L = 1/H_{\rr{ini}}$.  }
\label{fig:oneDplot}
\end{center}
\end{figure*}

In order to validate our codes, we have used the following six-point procedure:
\begin{enumerate}
\item \textit{Initial conditions:} code cross-check of initial $\phi$ values on the lattice between the two codes.
\item \textit{Initial conditions:} monitoring of the Hamiltonian constraint on the lattice, more precisely $\mathcal H^{\rr{rel.}} $ defined as the relative difference between geometric and matter terms in $\mathcal H$, see Eq.~(\ref{eq:Hconstrain}).   Testing the scaling of the Hamiltonian $L^2$ norm with the discretization step.
\item \textit{Homogeneous evolution:} simple homogeneous case and validation with the analytical FLRW expectation.
\item \textit{Inhomogeneous evolution:}  code cross-check.  Every $N_t$ time steps (typically $N_t$ goes from $100$ to $10^4$ depending on the lattice size and time step), export and cross-check of the dynamical quantities on some 1D or 2D lattice slices, or on the whole 3D lattice.
\item \textit{Inhomogeneous evolution:} at all time steps, monitoring of $\mathcal H^{\rr{rel.}} $ at several points and on average on the whole lattice.  \item \textit{Inhomogeneous evolution:} cross-check between the two codes of the evolution of the metric variables, density field and Hamiltonian constraint, at several points and after Riemannian averaging on the whole lattice.  Confronting the scaling of the Hamiltonian $L^2$ norm with the discretization time step to the order of the temporal scheme. 
\end{enumerate}

Because cross-checks between codes cannot fully guarantee the validity of the simulations, the above validation procedure includes several checks of the Hamiltonian constraint, which guarantees that our solution stays close to the one of general relativity.  We also tested the scaling of the Hamiltonian constraint with the number of lattice points and the discretization steps, which has to respect the order of the numerical schemes.  

On Fig.~\ref{fig:phiinitHrel} are shown typical initial values of $\phi$ and $\mathcal H^{\rr{rel.}}$ on the lattice, obtained with the two codes for a single two-dimensional inhomogeneity with a single mode in each direction $\texttt{x}$ and $\texttt{y}$ and vanishing phases.   The initial values of $\exp(\phi)$ obtained with the two codes agree at the $10^{-6}$ level.   The Hamiltonian constraint is satisfied at the $10^{-4}$ level when comparing geometric and matter terms of Eq.~(\ref{eq:Hconstrain}), a precision that is actually even improved for lower initial density contrasts.  This degree of accuracy can be reached typically as long as the physical size of the lattice is initially about the Hubble radius $H_{\rr{ini}}^{-1}$.

Regarding the time-evolution, the homogenous case reduces to the FLRW case with a very high degree of accuracy (at the $10^{-9}$ level), as expected.   In the inhomogeneous case, we have first cross-checked the evolution of the relevant dynamical quantities between the two codes.   Their evolution is represented on Fig.~\ref{fig:evolvalid}.  We found a very good agreement between the two codes (at most at the $10^{-4}$ level).  We considered evolutions for which $\mathcal H^{\rr{rel}}$ remains lower than the percent level.   An example of time evolution of $\mathcal H^{\rr{rel}}$ is shown on Fig.~\ref{fig:Hrel}.

\section{Simulations}  \label{sec:results}

In this section are presented the main results of our simulations:  a) for a single inhomogeneity in one dimension, b) for a single inhomogeneity in two and three dimensions, c) for inhomogeneities obtained from a sum of density modes with random phases, in three dimensions.  

\subsection{One-dimensional inhomogeneity}

The simplest considered case (beyond homogeneous simulations) is a single mode inhomogeneity, arbitrarily chosen to be along the \texttt{x} direction.  This case allows to understand more easily the evolution of the dynamical quantities $\rho, K, \phi, \bar \gamma_{ii}$ and $\bar A_{ii}$ (the off-diagonal components being vanishing, at the level of the numerical noise) and to identify which terms are at play in the evolution equations and in inducing the backreactions.

Fig.~\ref{fig:oneDplot} illustrates this evolution for a typical example with $\tilde \delta ( k_x = 2 \pi /L ) = 0.03 $ initially.   As expected, the inhomogeneity grows in time and becomes mildly non-linear at the end of the simulation.  One reaches a density contrast up to $\delta \rho / \langle \rho \rangle \simeq 0.4$ at the center of the over-dense region and down to $\delta \rho / \langle \rho \rangle \simeq -0.3$ within the under-density.   The over-dense regions thus becomes more quickly non-linear than the under-dense ones, as expected in structure formation.
Starting from a constant value, the extrinsic curvature $\vert K \vert$ hugs progressively a shape opposite to the density profile, indicating that the under-dense region experiences a higher expansion rate, and inversely.   This observation is confirmed in the profile of the conformal factor $\phi$ indicating that under-dense regions expand more than over-dense ones.   The shape of $\bar \gamma_{ii}$ indicates that within the under-dense region, the lattice cells squeeze progressively in the $\texttt{y}$ and $\texttt{z}$ directions orthogonal to the inhomogeneity and extend in the $\texttt{x}$ direction, in such a way that their volume scales like $\propto \exp(6 \phi)$  (let remind that the determinant of $\bar \gamma_{ij}$ is one by definition).   The opposite situation is observed in the over-dense region.   Such a shrinking is induced by negative (positive) values of $\bar A_{11}$ combined with positive (negative) values of $\bar A_{22}$ and $\bar A_{33} $ within the under-density (over-density).    Finally one can note that at the end of the simulation, the $A_{ii}$ develop some local features that are numerical artifacts, a sign that our results cannot be trusted anymore beyond this point.  This approximatively corresponds to the time beyond which the Hamiltonian constraint is no longer satisfied. 

\begin{figure}[]
\begin{center}
\includegraphics[width=75mm]{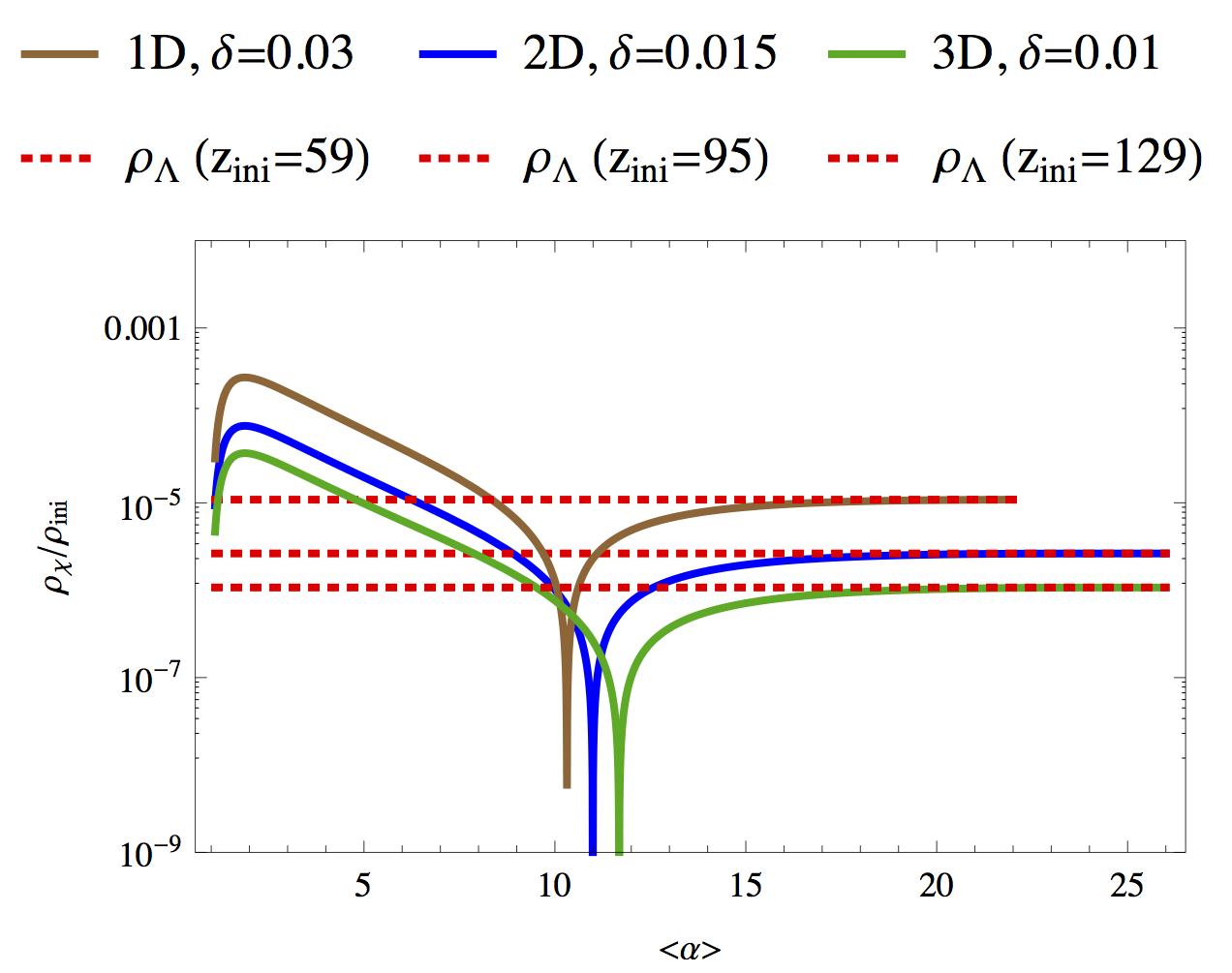} \\
\includegraphics[width=70mm]{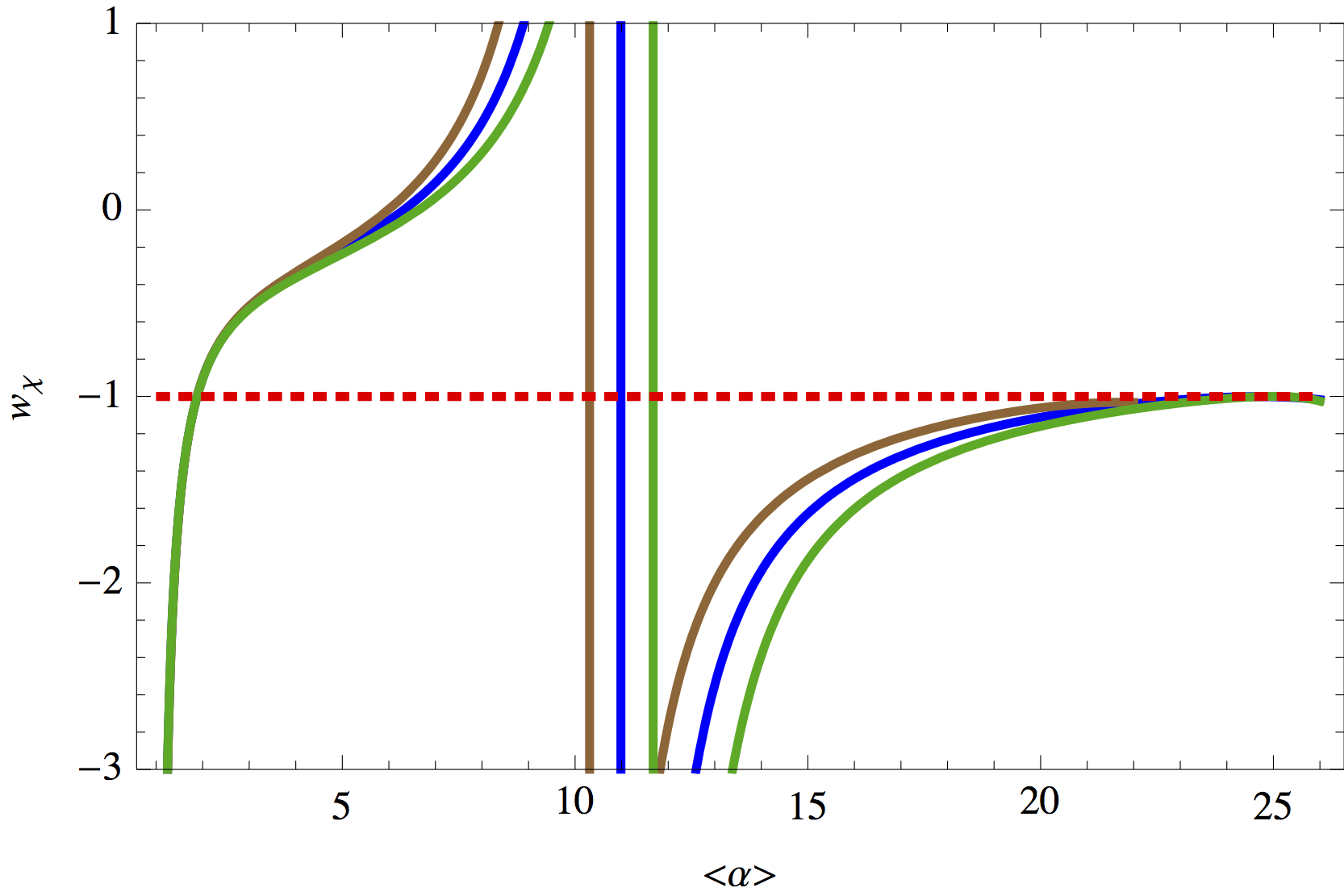} \\
\includegraphics[width=70mm]{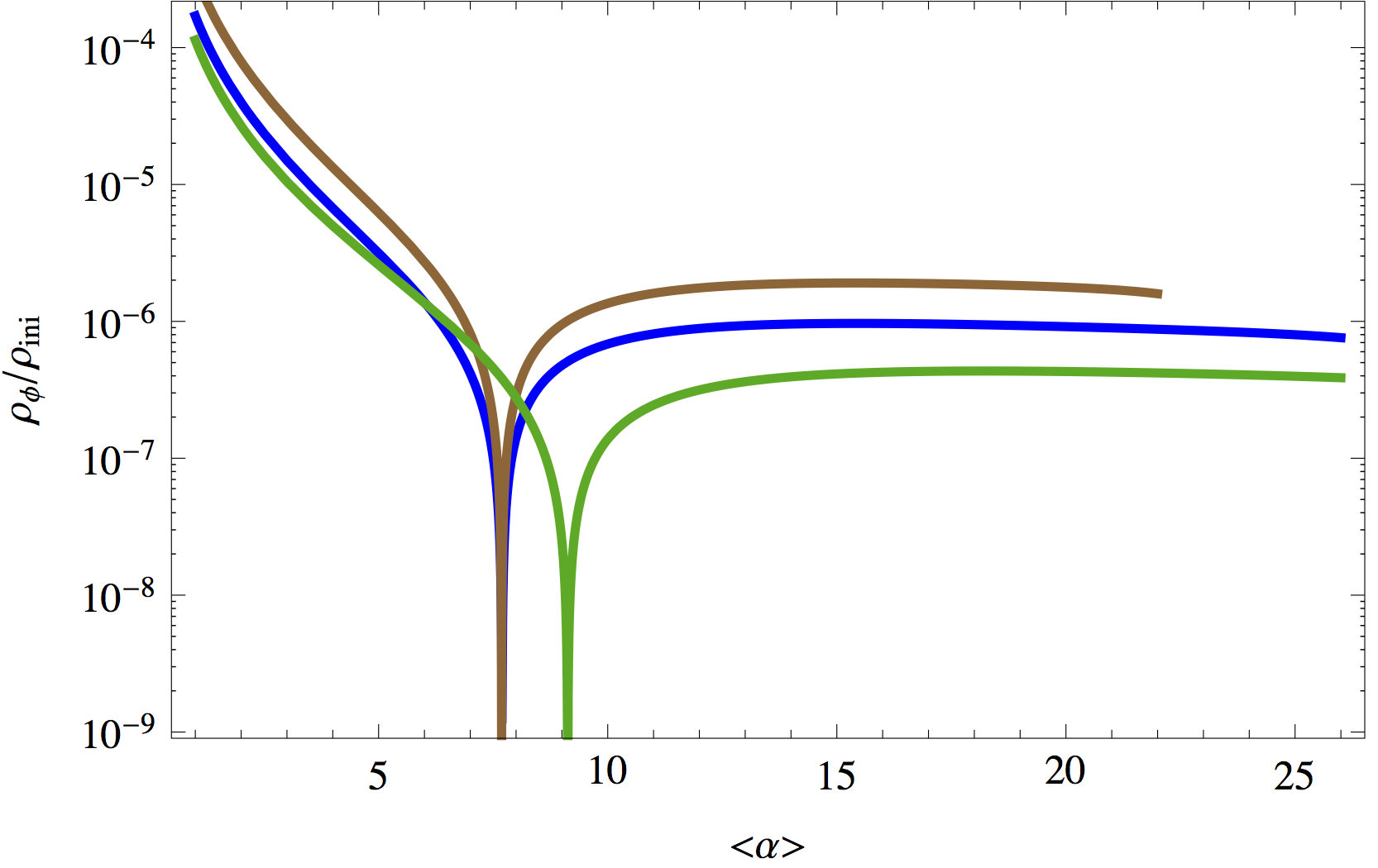} \\
\includegraphics[width=70mm]{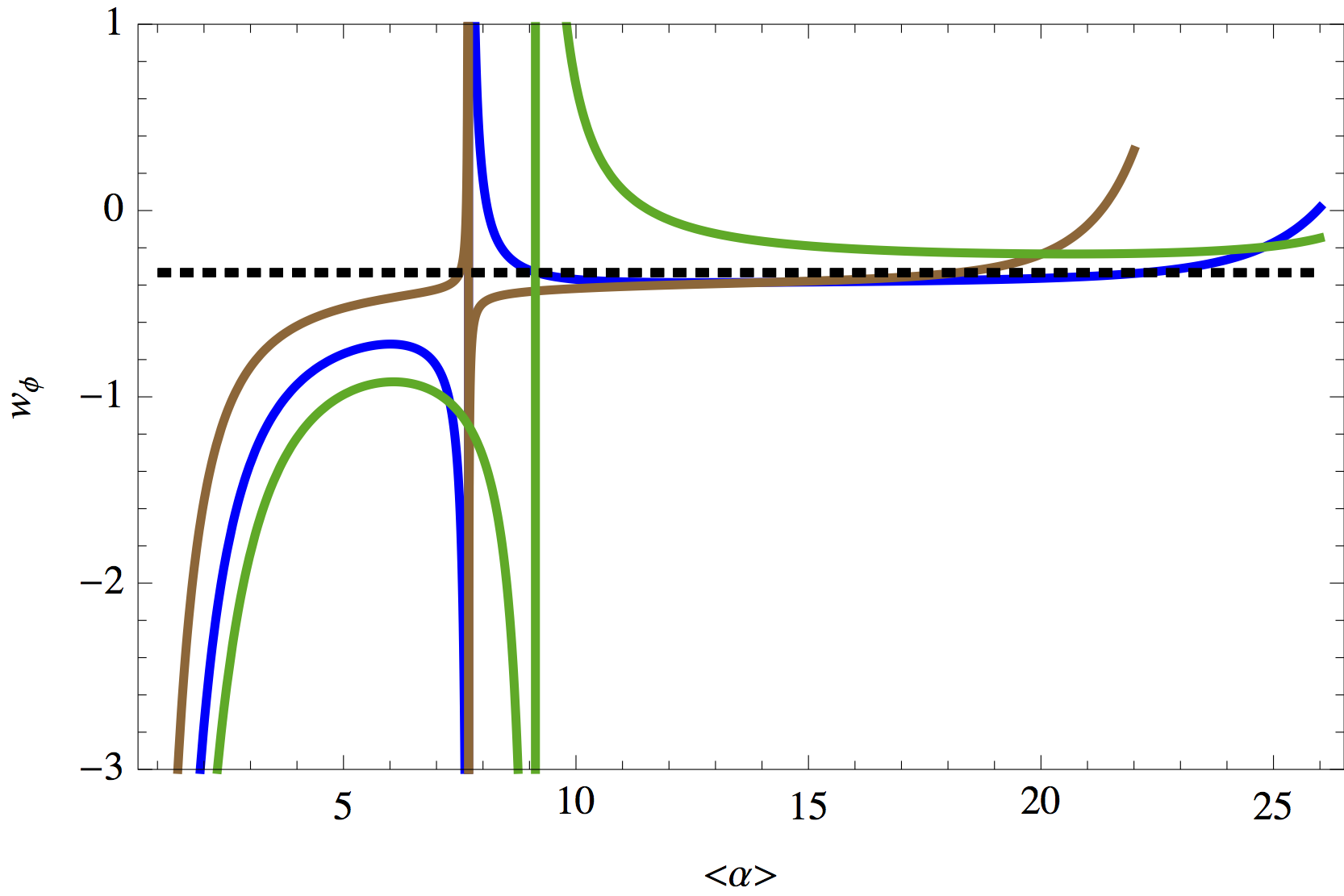} \\
\caption{Evolution, as a function of the averaged scale factor $\langle \alpha \rangle$, of the energy (top and third panel) and equation of state (second and fourth panels) associated to the two \textit{morphons} $\chi $ and $\varphi$, for a 1D (brown), 2D (blue) and 3D inhomogeneity (dark green) with respectively $\delta(k_x) = 0.03$, $\delta(k_{x,y} = 0.015 $ and $\delta(k_x,k_y,k_z) = 0.01 $.   The expected densities for a cosmological constant and the equation of states $w = -1$ and $w=-1/3$ are also displayed.     
 }
\label{fig:plotevola}
\end{center}
\end{figure}

In addition to spatial profiles, we have computed and plotted on Fig.~\ref{fig:plotevola} the evolution of the backreactions, seen in the FLRW picture as the the energy densities and equations of state of the two \textit{morphons}, $\rho_{\chi, \varphi}$ and $w_{\chi, \varphi}$.    The contribution from $\rho_{\chi} $ is first negative and exponentially decays with the scale factor $\langle \alpha \rangle$, until it becomes positive at $\langle \alpha \rangle \sim10$ and finally reaches a constant value that depends on how strong is the inhomogeneity initially.   At that time, $w_\chi$ goes from positive to negative values and then slowly increases to reach $w_\chi \approx -1$  at late times.   The \textit{morphon} $\chi$  therefore acts as a tiny cosmological constant ($\rho_{\chi} \sim 10^{-5} \rho_{\rr{ini}}$ for an initial density contrast $\delta \sim 10^{-2}$).   Its  contribution to the total density grows with time.  At the end of the simulation, one reaches $\langle \alpha \rangle \approx 20$ and $ \rho_{\chi} \sim 0.1 \langle \rho \rangle $, not enough to reach the matter-\textit{morphon} equality but enough to have a non-negligible impact on the apparent background expansion.   In order to interpret such a backreaction as Dark Energy, one would have to extrapolate $\rho_{\chi} $ to later times.  One would get $\Omega_\Lambda \simeq 0.7$ for initial conditions fixed at a redshift $z\sim 60$, for density contrasts $\delta \sim 10^{-2}$.  

The other \textit{morphon} first acts like a negative energy density that becomes then positive with $\rho_{\varphi} \sim 10^{-6} \rho_{\rr{ini}}$ and slowly decays with an equation of state close to $w_\varphi \approx -1/3$, i.e. it acts as a curvature-like fluid and decreases like $ \rho_\varphi \propto \langle \alpha \rangle^{-2}$.  This behavior can be explained if the backreaction term $\langle ^3 R \rangle_{\mathcal D} $ dominates over $Q_{\mathcal D}$ at late time  in Eq.~(\ref{eq:adotoaback}), which can effectively be interpreted as a curvature-like fluid.  

It is worth noticing that the purely relativistic terms involving non-vanishing $\bar A_{ii}$ in the BSSN equations cannot be neglected, but are not the only driver of the backreactions.   A similar level of backreactions is actually obtained if one neglects these terms, which also implies that one can identify the scale factors $\langle \alpha \rangle $ and $ \langle a \rangle$ (but not $a_{\mathcal D}$).   Within this approximation, each lattice cell behaves like a mini-FLRW Universe.  The backreaction effect is therefore somehow already captured by looking at the different expansion rates induced by the matter inhomogeneities.   Under this assumption it is also much easier to solve the BSSN equations that are not PDE's anymore but a system of coupled ODE's, to be solved at each lattice point.   But the approximation is not accurate and one should remind that it is is no longer valid when quantifying with some accuracy the level and the evolution of the backreactions.  

\subsection{Two and three-dimensional inhomogeneity}

The cases of a single two and three-dimensional inhomogeneity are rather similar to the one-dimensional case, as illustrated on Figs.~\ref{fig:oneDplot} and~\ref{fig:plotevola}, for density modes $\tilde \delta (k_{x,y} = 2 \pi / L) = 0.015 $ (2-dim) and $\tilde \delta (k_{x,y,z} = 2 \pi / L) = 0.01 $ (3-dim).   These values were chosen to keep constant the density contrast at the center of the inhomogeneity.   The displayed profiles are for positions $y = z = L/2$ on the lattice, and so they probe the evolution around the most under-dense region but not around the most over-dense one, which explains the differences observed in the $\bar \gamma_{ii} $  and $\bar A_{ii}$ profiles.  
The evolution of the backreactions are also very similar to the one-dimensional case and the same conclusions apply.   The $\chi$ field acts at late times as a tiny cosmological constant in the FLRW picture, whose value depends on how strong is the initial inhomogeneity, whereas $\varphi$ acts as a curvature-like fluid.   The $\chi$ field also exhibits a phantom-like equation of state $w_\chi < - 1$ before tending slowly $w_\chi \approx -1$.     Again, the $\bar A_{ij}$ driving the evolution of $\bar \gamma_{ij}  $  were found to not be the only driver of the backreactions, which already occur when their effect is neglected.  




\subsection{Multi-mode three-dimensional inhomogeneities}

\begin{figure*}
\begin{center}
\includegraphics[scale=0.55]{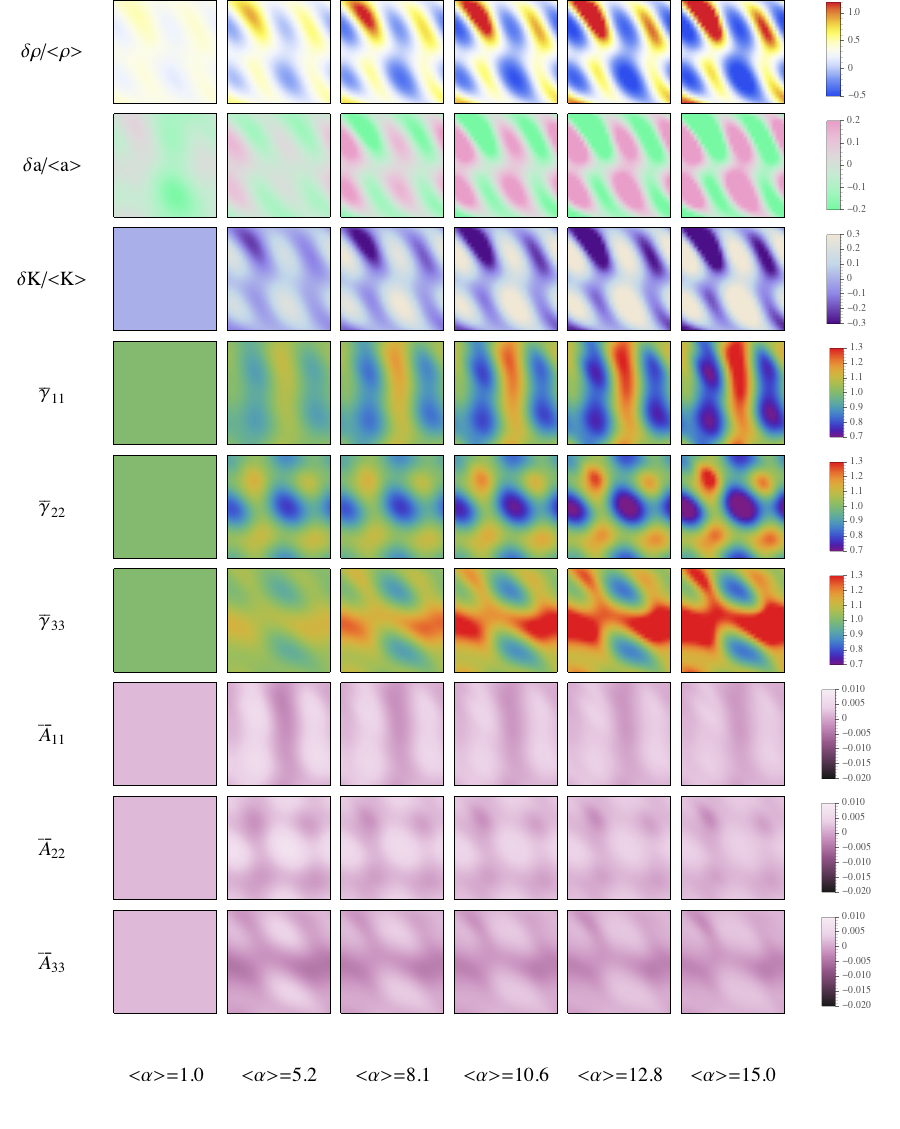} \\
\caption{Lattice slice showing the evolution (from left to right) of the dynamical variables $\rho$, $K$, $\phi$, $\bar \gamma_{ii} $ and $\bar A_{ii}$ (from top to bottom) for multi-mode 3D inhomogeneities with $\tilde \delta({\bf k}) = 0.01$ and $L = 1/H_{\rr{ini}}$.  }
\label{fig:plotarray}
\end{center}
\end{figure*}

In the real Universe, initial density fluctuations can be seen as a random superposition of gaussian fluctuations of different sizes and amplitudes, but exhibiting a nearly scale invariant power spectrum.  In order to approach this situation with our toy model, we have run simulations with initial density exhibiting multiple wavelength modes.  More precisely, we considered two modes along each spatial directions, $k_{x,y,z} = n \times 2 \pi / L$ with $n \leq  2$, i.e. 26 modes in total (not counting the homogeneous case  $k_{x,y,z} = 0$), with random phases.   Including higher modes is in principle possible but this would require to run larger simulations.   

Typical two-dimensional lattice profiles showing the evolution of the contrasts of $\rho$, $K$ and $a$ as well as the dynamical metric variables $\bar \gamma_{ii}$ and $\bar A_{ii}$  are displayed on Fig.~\ref{fig:plotarray}, for $\tilde \delta ({\bf k}) = 0.01 $.    Our simulations can proceed up to the time when the density contrast becomes non-linear, with maximal $\delta \rho / \langle \rho \rangle \simeq 1.2$ ($-0.4$ within the deepest under-density).    As it was observed for a single inhomogeneity, under-dense regions expand more than over-dense ones, as one can see on the $K$ contrast profiles.   Actually, at the end of the simulation, one can deduce from the $\delta a / \langle a \rangle$ profiles that the volume of the most under-dense lattice cells is about five times superior than for the most over-dense ones.   Driven by $\bar A_{ij}$, the $\bar \gamma_{ij}$ evolve such that the lattice cells shrink more in the direction of the lowest density gradients, as it was observed for a single inhomogeneity.   At the end of the simulation, similar numerical artifacts to the ones observed in the 1D case appear on the $\bar A_{ii}$, that we attribute to the numerical instability taking place when the Hamiltonian constraint is no longer satisfied.  Beyond that time, the simulation thus cannot be trusted anymore.
Similar behaviors have been obtained when lowering the initial mode amplitude, down to $\tilde \delta ({\bf k})  = 10^{-4}$, the only noticeable differences being that it takes obviously more time for the density contrasts to deepen and become non-lienar, and so for the inhomogeneities to induce a significant backreaction.  

\begin{figure}
\begin{center}	
\includegraphics[width=85mm]{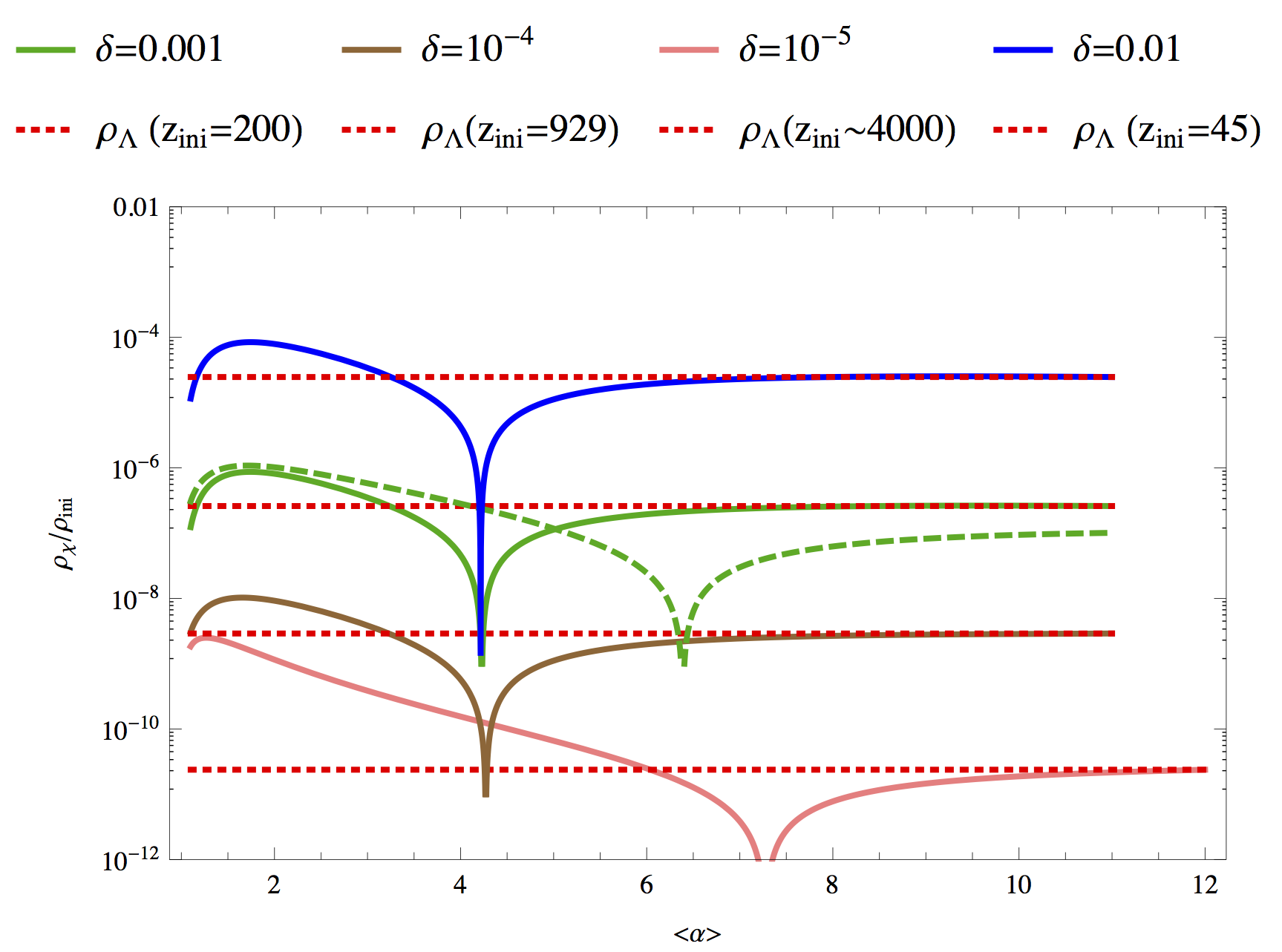} \\
\includegraphics[width=75mm]{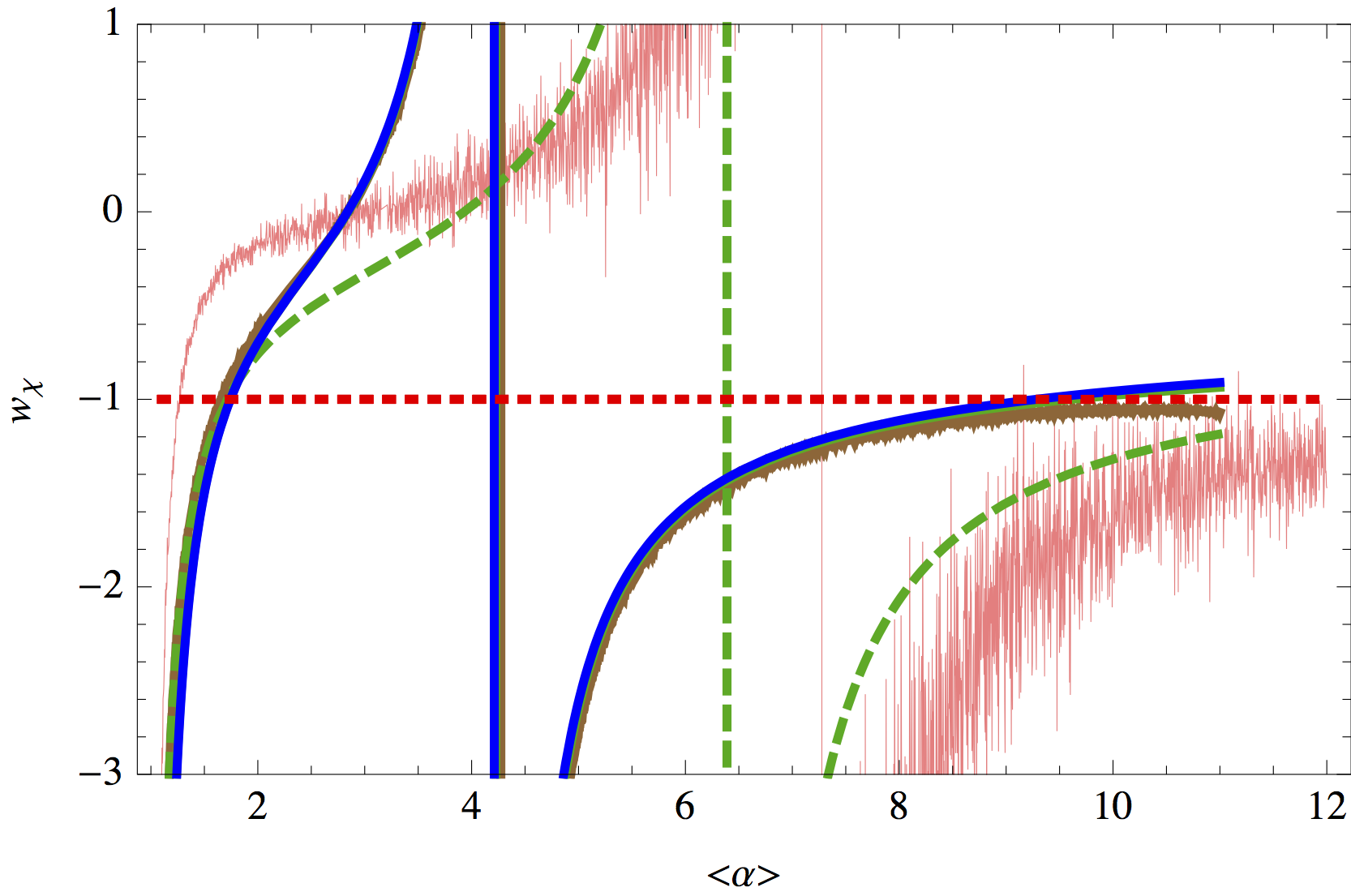} \\
\includegraphics[width=75mm]{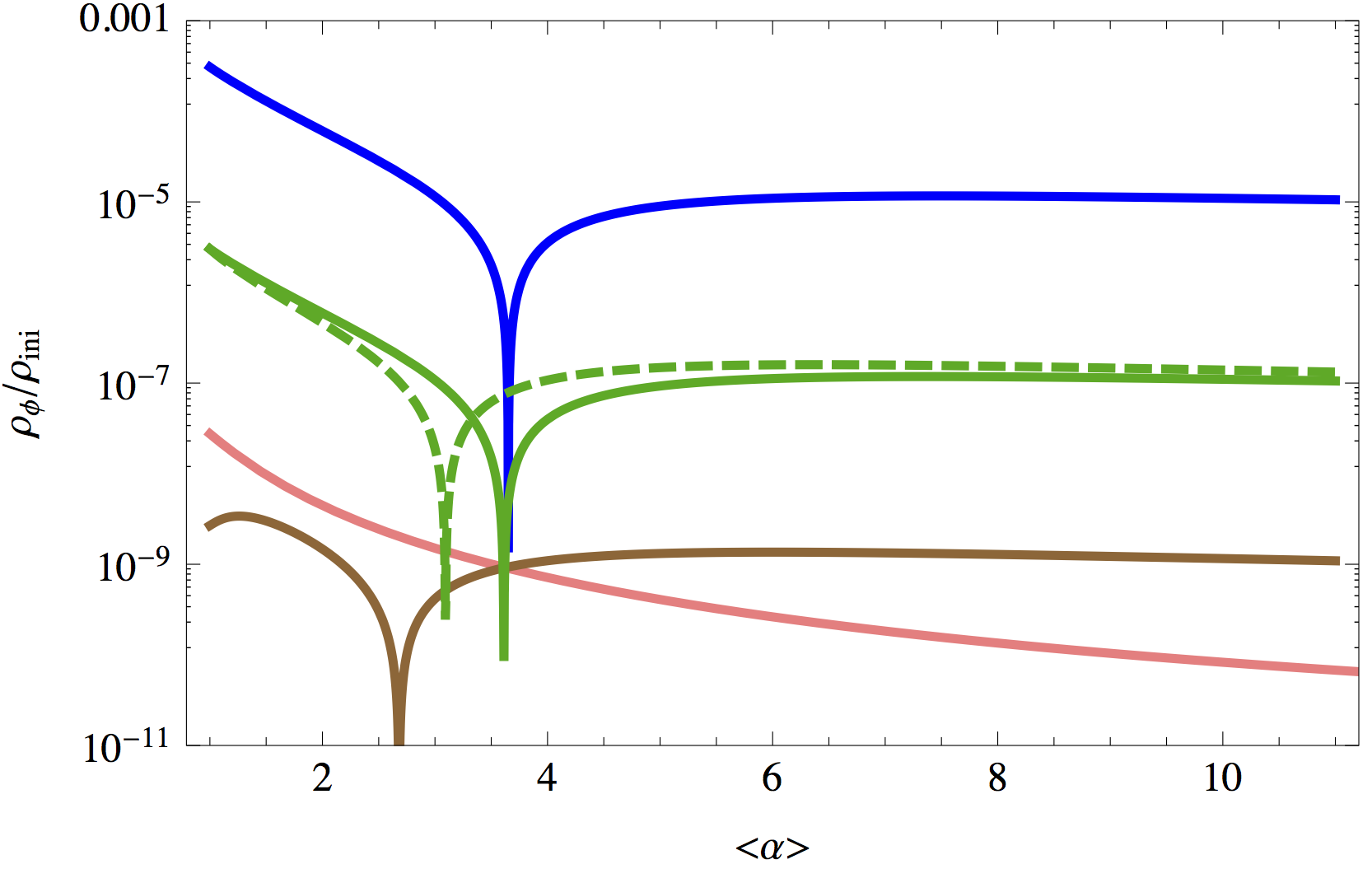} \\
\includegraphics[width=75mm]{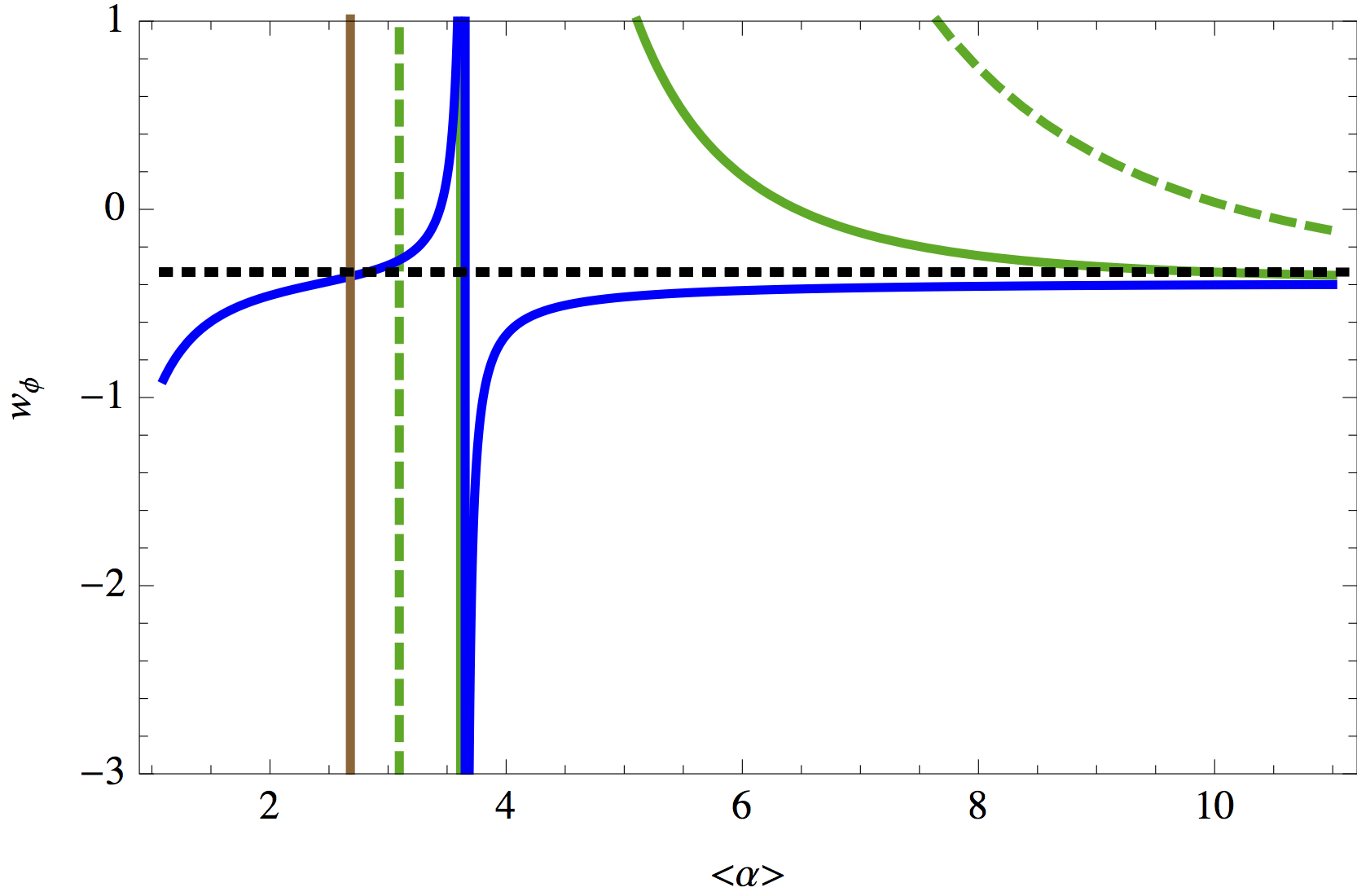} \\
\caption{Evolution of the energy (top and third panel) and equation of state (second and fourth panels) associated to the two \textit{morphons} $\chi $ and $\varphi$, for multi-mode 3D inhomogeneities with initial amplitude in the range $10^{-2} \leq \tilde \delta({\bf k}) \leq 10^{-5}$.  The dashed green curves correspond to the case $\bar A_{ij} = 0$, i.e. no other local general relativistic effect than lattice cell expansion as mini-FLRW universes.          
 }
\label{fig:plotevol3Db}
\end{center}
\end{figure}

In the FLRW picture, we observe a similar behavior in the evolution of the two \textit{morphons}, represented on Fig.~\ref{fig:plotevol3Db}.  On one side, $\rho_\chi$ is first negative but at some point it becomes positive and its equation of state exhibits an  asymptote.  Then $\rho_\chi$ increases slowly toward a constant value that can be interpreted as a tiny cosmological constant since its equation of state tends towards $w_\chi \approx -1$.   On the other side, $\varphi$ behaves at late time as a curvature-like fluid with an equation of sate tending towards $w_\varphi \approx -1/3$.   This behavior is observed whatever is the initial amplitude of the inhomogeneities, within the range $10^{-4} \lesssim \tilde \delta ({\bf k}) \lesssim 10^{-2}$.   For even lower values, one can observe that the evolution of $w_\chi$ becomes strongly contaminated by the numerical noise, and therefore it is impossible to conclude on the level and evolution of the backreactions.   Nevertheless the general tendency is a similar behavior for the $\chi$ field, but with a $\rho_{\chi}$ so tiny that one would have to extrapolate our results between redshifts $z\sim 300 $ and today, as well as to assume that $w_\chi \approx -1$ all during that period, in order to interpret this effect as Dark Energy, which is hazardous.  On the other hand, $w_\phi$ is found to take very large values, far from $-1/3$.   We therefore conclude that improvements of the codes combined with larger simulations will be required in order to probe this potentially interesting regime.   But our analysis does not rule out the possibility that important backreactions take place today, induced by $\sim 10^{-5}$ density perturbations, as expected in the standard cosmological scenario.

Finally, one should comment on the importance of the purely relativistic effects from $\bar A_{ij}$ terms sourcing the conformally rescaled metric $\bar \gamma_{ij}$ as well as the extrinsic curvature $K$.   When the  $\bar A_{ij}$ are artificially set to zero, the qualitative evolution of the backreactions is still present, and especially the late-time CC-like behavior of $\chi$.   Nevertheless, as shown on Fig.~\ref{fig:plotevol3Db}, not including this effect implies that the transition towards a negative $w_\chi$ occurs at later time, and that $\rho_{\chi} $ is about one order of magnitude lower than in the fully relativistic case.   

\section{Backreactions \\†mimicking dark energy} \label{sec:backreactions}

The principal result of the simulations described in the previous section is that the \textit{morphon}  field $\chi$ can mimic a Dark Energy component, for initial density fluctuations within the range $10^{-4} \lesssim \tilde \delta(k) \lesssim 10^{-2}$ and of the size of the Hubble radius.   Even if the simulations are not stable enough to reach the matter-\textit{morphon} equality, we have shown that the density associated to $\chi$ tends to a constant value, with an equation of state tending to $w_\chi \approx -1$.  It accounts for a non-negligible part (up to about $10\%$) of the total density at the end of the simulations, enough to have already an important impact on the expansion.  When this behavior is extrapolated at later times, it is expected to become dominant, and we have inferred the initial redshift of the simulations so that the backreactions would lead to Dark Energy with $\Omega_{\rr{DE} } = 0.7$ today.    We found initial redshifts $z_{\rr{ini}} \approx 60 $ for $\delta(k) \approx 10^{-2}$, going up to  $z_{\rr{ini}} \approx 1000 $ for $\delta(k) \approx 10^{-4}$.    For lower inhomogeneities, the backreactions are contaminated by numerical noise and thus our results are inconclusive even if one still observes the tendency to reach a CC-like regime for $\chi$.  

The validity of the extrapolation is not entirely guaranteed, but it is supported by the fact that in absence of the purely relativistic effects induced by $A_{ij}$ in the evolution equations, $\rho_{\chi}$ is found to stay constant with $w_\chi \approx -1$ over the required period of time.  The observed freeze-out of $A_{ij}$, induced by the decay of $K$ and $R_{ij}^{\rr{TF}}$ in Eq.~(\ref{eq:Aijevol}), also supports the hypothesis that $\chi$ continues to act as a CC as long as no singularity is developed, which in the Universe is prevented by the virialization of structures. 

Interestingly, in a scenario with $\delta \sim 10^{-2}†$ initially, we predict a phantom-like equation of state $w_\chi \lesssim -1$ at redshifts $z \gtrsim 5$ ($z \gtrsim 3 $ for a single inhomogeneity) .   A prediction that could eventually be tested by future 21cm intensity mapping experiments like the Square Kilometre Array, or even by future LSS surveys like Euclid.  The recent analysis of~\cite{Hee:2016nho} claiming that combined CMB and LSS data slightly prefer a super-negative equation of state $w \lesssim -1$ at redshifts $z \gtrsim 3$, could be a hint in favor of such a scenario.  
Finally, one should comment on the curvature-like effect of the second \textit{morphon} $\varphi$.   The observations do not prevent such a backreaction that actually would still be subdominant today and would stay within the present bounds on $\Omega_K$.   

\section{Conclusion and discussion}  \label{sec:ccl}
 
The backreactions induced by cosmic inhomogeneities on the global expansion dynamics have been evaluated for a toy model Universe filled entirely with a pressure-less matter fluid  assimilated to the Dark Matter, and initial density contrasts in the range $10^{-4} - 10^{-2}$ on scales initially crossing the Hubble radius.  For this purpose, we have developed and used two numerical relativity codes based on the 3+1 BSSN formalism, gathered in the  \textit{Inhomogeneous Cosmology and Relativistic Universe Simulations} (\texttt{ICARUS}) package.  Given some initial density distribution, the codes first solve the Hamiltonian constraint on a uniform real-space lattice with periodic boundary conditions, then evolve in time the BSSN equations and finally proceed to the Riemannian averaging of the relevant quantities.

The main objective of this paper was to identify the required properties for the initial inhomogeneities to induce an important backreaction on the expansion when the observations are interpreted in the FLRW picture.   Besides the characterization of the backreactions, a second objective was to determine whether such a backreaction could eventually mimic a Dark Energy component, for this particular toy mode.   

 Backreactions of two origins have been distinguished and interpreted  as the effect of two \textit{morphon} scalar fields.  Backreactions of two kinds have been distinguished:  i) the ones affecting the Friedmann-Lema\^itre equations given the averaged density, ii) the ones modifying the time evolution of the averaged energy density itself.   The latter effect was overlooked in previous works and actually plays a crucial role.   Characterizing the backreactions also needs to distinguish between three non-equivalent \textit{definitions} of the scale factor in an inhomogeneous Universe:  i) $a_{\mathcal D} $ that rescales the total domain volume like $a_{\mathcal D}^3$, ii) $\langle a \rangle$ that is the averaged (in a Riemannian way) of the local scale factor (i.e. the factor scaling the volume of individual lattice cells in the simulations), iii) $\langle \alpha \rangle $ that is the average of the local factor rescaling the proper lengths of individual cells in a given direction.  The latter is the one used for redshift and distance measurements, i.e. the one inferred by observations.  

Whereas the backreactions of the first kind simply behave like a curvature fluid that is damped with the expansion and falls below the current observational limits on $\Omega_K$, the ones of the second kind act in the FLRW picture as a scalar field with an equation of state tending to $w \approx - 1$, after a regime where $w < -1$,  which does not imply any theoretical issue since the underlying theory is General Relativity only.   Therefore this \textit{morphon} acts as a tiny cosmological constant, that could eventually mimic Dark Energy and the cosmic acceleration of the expansion.  Our simulations can probe the evolution of inhomogeneities up to the time when the backreactions contribute to about $\Omega_{\rr{DE}} \simeq 0.1$.   When this behavior is extrapolated until today, one can find the redshift and scale of the initial density fluctuations leading to $ \Omega_{\rr{DE}} \simeq 0.7$ today.  For instance, inhomogeneities with a density contrast of $\sim 10^{-2} $ should enter inside the Hubble radius at $z \sim 45$ and would correspond to gigaparsec  fluctuations today, with a transition from a phantom-like to a cosmological constant-like equation of state occurring at redshifts $z \sim 4-7$, depending on the initial density pattern.  
 Testing the validity of this extrapolation will require larger simulations combined with further improvements of the code stability, over a longer period of time.    Nevertheless we gave some qualitative arguments supporting its validity, e.g. the fact that the simulations reach a regime where the global backreaction and the general relativistic effects, except the expansion of local regions behaving like mini-FLRW universes, are frozen.  

A super-negative equation of state for Dark Energy at high redshifts is a general prediction of the proposed scenario, which will be tested by future 21cm and LSS experiments.  Recent works based on  CMB and LSS data actually already slightly favored this case against a cosmological constant~\cite{Hee:2016nho}.  It is also possible that such a scenario with large scale inhomogeneities would relieve the tension observed in $H_0$ measurements, in a similar way than the one proposed e.g. in~\cite{Biswas:2010xm}.    The model presented in this paper is however still at the level of a toy model: it does not include baryons and radiation, virialization of structures, and implements simple inhomogeneity patterns on a restricted range of scales.   Such matter fluctuations on comoving scales $k\sim 10^{-4}$ Mpc$^{-1}$ are probably in some tension with CMB anisotropies measured by Planck~\cite{Ade:2015lrj}.   A more precise investigation would be needed to study the viability of the scenario.   One can notice that larger scales, $k \lesssim 10^{-4}$ Mpc$^{-1}$, are very poorly constrained with CMB observations, which potentially leaves some range for our model to be viable.  Also, our analysis does not exclude the possibility of backreactions arising from $10^{-5}$ fluctuations crossing the Hubble radius close to the matter-radiation equality.    

The observations of troubling structures on gigaparsec scales, such as the cold spot and some super-voids~\cite{Finelli:2014yha,Kovacs:2015hew} invoking density contrasts that are in strong tension with the statistical expectations of the standard cosmological model, but similar to the ones obtained in our simulations, could be a hint in favor of such a scenario. 
 
 Mimicking  Dark Energy with the backreactions  from matter inhomogeneities induced by some power spectrum enhancement on the largest cosmological scales is a new mechanism  that could send the fine-tuning and coincidence issues of Dark Energy back to the origin of the enhancement in the primordial power spectrum.  This kind of behavior could be produced in some inflation models (see e.g.~\cite{Clesse:2013jra} for an example in the context of hybrid inflation).  
  
On the point of view of code development, several perspectives are envisaged, such as including the implementation of other gauge choices that could allow to run simulations deeper in the non-linear regime or to include several fluids.  More performant integration schemes (e.g. a predictor corrector build on a \texttt{rk4} method, symplectic integrator) and new methods for solving the problem of initial conditions could be also implemented.  A version of the code dedicated to scalar field cosmology is under development and could be used for various problems such as the inhomogeneous initial conditions of inflation, the tachyonic preheating, or the dynamics of  quintessence field fluctuations.    
Refining the description and the evolution of backreactions would also require to include several effects:  a more precise implementation of the initial density fluctuations, in term of the primordial power spectrum, the effect of light propagation through the inhomogeneous Universe and the signatures of the specific local inhomogeneous environment of the observer.   

Our work therefore contributes to pave to road of large cosmological simulations in numerical relativity, that would allow to study and reveal all the general relativistic effects on the cosmological dynamics, including the level of backreactions and their potentially observable signatures.

\section*{Acknowledgments}  
The authors warmly thank Julien Larena, Julien Lesgourgues and Jeremy Reckier for useful discussions.  Large simulations were realized thanks to the High-Performance-Computing facilities of the RWTH Aachen University.  

\bibliography{biblio}

\end{document}